\documentclass[useAMS,usenatbib]{mn2e}
\usepackage{graphicx}

\newcommand{\ms}{{\rm M_\odot}}

\newcommand{\teff}{$T_{\rm eff}$}
\title[Evolutionary Stellar Population Synthesis at High Spectral Resolution: Optical wavelengths]
{Evolutionary Stellar Population Synthesis at High Spectral Resolution: Optical Wavelengths}
\author[R.M. Gonz\'alez Delgado et al.]
{R.M. Gonz\'alez Delgado$^1$\thanks{E-mail:
rosa@iaa.es (RGD); mcs@iaa.es (MC); martins@stsci.edu (LPM); leitherer@stsci.edu (CL); 
yeti@hs.uni-hamburg.de (PH)} 
, M. Cervi\~no$^1$, L.P. Martins$^2$, C. Leitherer$^2$, P.H. Hauschildt$^3$\\
$^{1}$Instituto de Astrof\'\i sica de Andaluc\'\i a (CSIC), Apdo. 3004, 18080 Granada, Spain\\
$^{2}$Space Telescope Science Institute, 3700 San Martin Drive, Baltimore, MD 21218 \\
$^{3}$Hamburger Sternwarte, Gojenbergsweg 112, 21029 Hamburg, Germany}

\begin{document}

\date{Accepted 2004 November. Received 2004 May; in original form 2004 May}

\pagerange{\pageref{firstpage}--\pageref{lastpage}} \pubyear{2005}

\maketitle

\label{firstpage}

\begin{abstract}

We present the single stellar population (SSP) synthesis results of our new synthetic
stellar atmosphere models library with a spectral
sampling of 0.3 \AA, covering the wavelength range from 3000 \AA\ to 7000
\AA\ for a wide range of metallicities (twice solar, solar, half solar and 1/10 solar).  
The stellar library is composed of
1650 spectra computed with the latest improvements in stellar
atmospheres. In particular, it incorporates non-LTE line-blanketed models for
hot ($T_{\rm eff}\geq27500$ K), and LTE line-blanketed models (Phoenix) for 
cool ($3000\leq T_{\rm eff}\leq4500$ K) stars. Because of the high spectral resolution of this library,
evolutionary synthesis models can be used to predict the strength of numerous 
weak absorption lines and the evolution of the profiles of the strongest lines 
over a wide range of ages. The SSP results have been calculated for
ages 1 Myr to 17 Gyr using the stellar evolutionary tracks provided
by the Geneva and Padova groups. For young stellar populations, 
our results have a very detailed coverage of high-temperature stars with similar results 
for the Padova and Geneva
isochrones. For intermediate and old stellar populations, our results, once
degraded to a lower resolution, are similar to the ones obtained by other
groups (limitations imposed by the stellar evolutionary physics notwidthstanding).
The limitations and advantages of our models for the analysis of integrated 
populations are described. The full set of the stellar library and the evolutionary
models are available for retrieval at the websites {\tt http://www.iaa.csic.es/$\sim$rosa} and
{\tt http://www.iaa.csic.es/$\sim$mcs/Sed@}, or on request from the first two authors.

\end{abstract}


\section{Introduction}

Evolutionary synthesis, introduced by
\cite{Tin68}, is a powerful tool for interpreting the integrated light of a stellar
population. This technique makes a prediction for the properties of a
stellar population taking as a free parameter the star formation rate, the
initial mass function (IMF) and the metallicity.  During the past decade,
evolutionary synthesis models have been developed and improved
significantly, using the newest and more advanced stellar evolutionary
tracks and stellar libraries
\cite[e.g.][]{CMH94,FRV97,SB99,Vaz99,Buz02a,galev02,BC03}.
In addition, they have been extended to new
wavelength domains \cite[e.g.][ for $\gamma$ and X-rays respectively]{Cetal00,CMHK02}. 
However, it is also well known that the power of
this technique is very dependent on the reliability of the stellar models
and libraries used as ingredients \cite[][and references therein]{LFvAH96},
as well as on the size of the system to which the models are applied
\cite[][]{CLC00,Bru00,Cervetal01,Cetal02,CL04,GGS04}. One important question related
to this technique is whether it is actually up to the challenge of
interpreting multi-wavelength data to infer the properties of stellar populations
of very different ages and metallicities \cite[e.g.][]{Orietal99,MO03}.

In recent years, numerous observational surveys have been undertaken to
measure the redshift evolution of the stellar properties of distant
galaxies, with the goal of understanding the formation and evolution
of galaxies. Spectral observations of intermediate and high redshift
galaxies have shown a large variety of stellar populations.  The integrated
light of these galaxies has been analyzed with evolutionary synthesis
models to derive ages, masses and metallicities. At
ultraviolet (UV) wavelengths, many of these galaxies show the footprints of recent bursts
of star formation. They are classified as star forming galaxies
\cite[][]{steidel}. The UV spectra of these galaxies, which are very similar to
local starbursts, are dominated by resonance lines formed in the
interstellar medium and/or in the winds of massive stars \cite[][]{pettini}. Observations
at rest-frame optical wavelengths of these high redshift galaxies (z$\geq$
2) are scarce due to the lack of non-cryogenic near-infrared (NIR) multi-object
spectrographs in very large telescopes.  H$\alpha$ nebular emission has
been detected in a few of these galaxies. However, as in nearby star
forming galaxies \cite[][]{Brietal03}, the higher order Balmer
lines appear in absorption, or in emission with strong absorption
wings. Thus, to estimate the star formation rate in these galaxies requires
a careful fit of the stellar continuum and subtraction of the absorption
component of the Balmer lines. Evolutionary synthesis models at
intermediate or at high spectral resolution are needed to perform this fit.

With this motivation, Gonz\'alez Delgado, Leitherer, \& Heckman (1999, hereafter GLH99)
built synthetic spectra of H Balmer and He
absorption lines of starbursts and post-starburst galaxies, 
using as input to the evolutionary code
Starburst99 \cite[][]{SB99} a stellar library of synthetic spectra, with a
sampling of 0.3 \AA, effective temperature from 50000 to 4000 K, and
gravity in the range 0.0 $\leq$log g$\leq$ 5.0 at solar metallicity
\cite[][hereafter GL99]{GL99}. The line profiles were computed using a set
of programs developed by Hubeny and collaborators \cite[][]{Hub88,HLJ95}. Due
to computational restrictions, we only computed small spectral ranges around the most
important Balmer and He{\sc i} lines.

Many beautiful data sets on starbursts and galaxies with active galactic 
nuclei (AGN) have been  
obtained at intermediate spectral resolution in the last few years to date the stellar population
\cite[]{GHL01,Cid01,Are, K03a,K03b,K03c,Cid04,G04,Tad}. 
These results have motivated us to build a much improved stellar library, both in the
mapping of the (\teff, log $g$) plane and in the complete coverage of the
optical spectral range at high spectral resolution. With this new library
it is possible to predict the spectral evolution not only of young
starbursts but also the properties of old stellar systems. Intermediate or
high spectral resolution models are also required to constrain the stellar
population in ellipticals, and to break the age-metallicity degeneracy
\cite[]{VazA99}.  With this motivation, \cite{BC03} presented
models, computed with their code {\sc galaxev}, at intermediate resolution (3
\AA) covering a large range in ages (0.1 Myr to 20 Ga), and wavelengths
(3200--9500 \AA) for several metallicities. {\sc galaxev} models are in some
aspects similar to those presented in this paper, but they have much lower
spectral resolution, different coverage of the (\teff, log$\,g$)
plane and, most importantly, they use empirical stellar spectra. 
A comparison of the two sets of models is presented Section 5. 

This paper is organized as follows. In Section 2, we briefly discuss the
stellar library. A more detailed description is given by
\cite{paperI}.  Section 3 describes the main stellar
ingredients, the assumptions, and the computational techniques of the
synthesis models. Section 4 presents the results and a discussion of the spectral
evolution of stellar populations, at several metallicities, which evolve
according to the Geneva or Padova isochrones.  The models are compared with
previous work in Section 5, and with observations of star clusters in the Large and Small Magellanic Clouds 
in Section 6. We discuss the limitation of the models in the red supergiant phase 
in Section 7. The summary and conclusions are in Section 8.

\section{Stellar Library}

This stellar library is an extension of the library built by
GL99. Here, we present a brief description. A full discussion is
given in \cite{paperI}.

The grid includes the synthetic stellar spectra from 3000 to 7000 \AA\ with 
a final spectral sampling of 0.3 \AA. The spectra span a range of effective 
temperature from 3000 to 55000 K, with variable steps from 500 to 2500 K, 
and a surface gravity log$\,g=-0.5$ to 5.5 with dex steps of 0.25 and 0.5.
For each temperature, the minimum gravity is set by the Eddington limit.
The library covers several metallicities: twice solar, solar, half and 1/10 solar. 
We assume solar abundance {\em ratios} for all the elements, and a helium 
abundance of He/H=0.1 by number. In addition to the spectral and metallicity coverage, 
this new library presents the following improvements with respect to GL99:

\begin{enumerate}
\item{ {\it Hot stars}.  Line-blanketed, non-LTE, plane-parallel,
hydrostatic atmospheres are used to compute the stellar spectra of O-type
stars.  The spectra are from the grid of \cite{LH03}\footnote{Available
at {\tt http://tlusty.gsfc.nasa.gov}}. This grid has 12 effective
temperatures, $27500\leq T_{\rm eff}\leq55000$ K, with 2500 K steps, and
eight values of the surface gravity, $3.0\leq{\rm log}\,g\leq4.75$, with 0.25
dex steps. These models assume a microturbulence velocity of 10 km
s$^{-1}$.  }
\item{ {\it Cool stars}.  Line-blanketed, LTE, spherical
atmospheres are computed with the PHOENIX code 
\cite[]{HB99,AHA01} 
for stars with $3000\leq T_{\rm eff}\leq4500$ K, with steps of 500 K. The grid covers
13 values of surface gravity, $-0.5\leq{\rm log}\,g\leq5.5$, with 0.5 dex
steps. The models assume a mass of 1 M$\odot$, a mixing-length equal to 2.0, and a
microturbulence velocity of 2 km s$^{-1}$. } 

\item{ {\it Intermediate temperature stars}. 
As in GL99, the synthetic spectra for stars with $8000\leq T_{\rm
eff}\leq27000$ K are obtained with Kurucz LTE atmospheres\footnote{Available at 
{\tt http://kurucz.harvard.edu/}} \cite[][]{Kurucz} and the program
SYNSPEC\footnote{Available at http://tlusty.gsfc.nasa.gov}\cite[][]{HLJ95}, except that for stars with
$4750\leq T_{\rm eff}\leq7750$ K, we use the program
SPECTRUM\footnote{Available at \\ {\tt http://www1.appstate.edu/dept/physics/spectrum/spectrum.html}} 
\cite[][]{spectrum}, together with Kurucz
atmospheres. Differences found between the metallic line strengths and
molecular bands using SPECTRUM and SYNSPEC are discussed in \cite{paperI}.}
\end{enumerate}
 
In all the cases, the maximum distance between two neighboring frequency
points for evaluating the spectrum is 0.01 \AA. However, the spectra are
degraded to have a final resolution of 0.3 \AA\ by performing a rotational and
instrumental convolution for each spectrum.
 
The library contains 415, 409, 416 and 411 spectra for Z=0.002, 0.010,
0.020 (solar) and 0.040, respectively. The typical grid coverage is
illustrated in Figure 1. Isochrones obtained with the evolutionary tracks
from the Geneva and Padova groups (cf. Section 3) have been over-plotted.
The figure illustrates that this library constitutes a homogeneous set of
stellar spectra covering a dense grid of fundamental parameters, including
a significant number of models that 
reduce the problems of the {\it close model assignation}, assumed in
most synthesis codes \cite[see, however, ][for a different
approach]{Lucetal04}. This library also presents some improvements with
respect to STELIB\footnote{Available at {\tt http://webast.ast.obs-mip.fr/stelib/}} \cite[]{LeBetal03}, the
stellar library used by {\sc galaxev}; in particular, it includes models
with $T_{\rm eff}\geq25000$ K, and $T_{\rm eff}\leq4000$ K and ${\rm
log}\,g\leq1.0$, which allow the prediction of the
spectral evolution of the stellar populations of both very young starbursts
and old systems. Furthermore, the high spectral resolution of the
library allows us to predict the evolution of the He{\sc i} and He{\sc ii}
lines, and to estimate the absorption correction to the nebular He emission
lines. This correction is very important for an accurate determination of
primordial He abundance using nebular emission lines of 
star forming galaxies.

The full set of the stellar library is available for retrieval at the website 
(http://www.iaa.csic.es/$\sim$rosa) or on request from the authors.

\begin{figure*}

\hbox{
{\rotatebox{270}{\includegraphics[width=6.5cm]{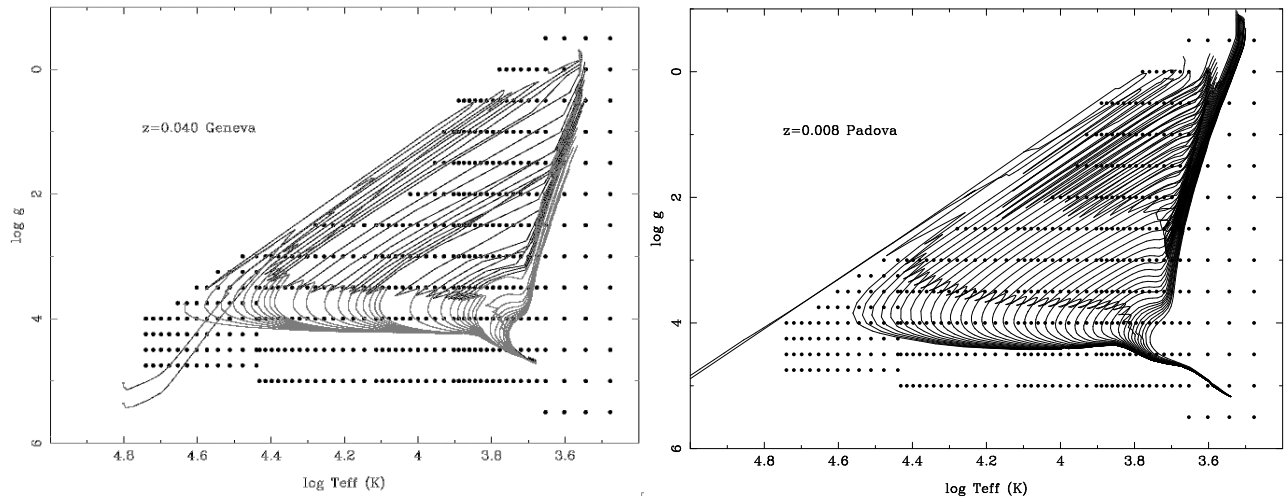}}}
{\rotatebox{270}{\includegraphics[width=6.5cm]{rgdf1b.eps}}}}

 \caption{HR diagram: Points represent the stellar library grid.
 Lines are isochrones of stellar populations that evolve following the Geneva (left)
and the Padova (right) evolutionary tracks at metallicities Z=0.040 and Z=0.008, 
respectively. The isochrone ages are from 1 Myr to 10 Gyr for the Geneva and 
4 Myr to 17 Gyr for the Padova models.}

\end{figure*}

\section{Description of the inputs}
%:

The library has been implemented in Starburst99 and Sed@\footnote{Sed@ is a
synthesis code included in the {\it Legacy Tool project} of the {\it
Violent Star Formation} European Network. The code is written in ANSI C
under GNU Public License and, currently, is coordinated by M. Cervi\~no; the code and their results
must be referred to solely by its documentation, {\it Sed@ Reference Manual}
(in preparation), its web address and the citations in the headers of the resulting files. More information can be found at
{\tt http://www.iaa.csic.es/$\sim$mcs/Sed@}}. There are no significant differences between the results obtained
with the two codes when models with the same ingredients (evolutionary
tracks, IMF, ....) are compared.  Here, we describe only the assumptions
and prescriptions related to Sed@ models. 

There is, however, a principal difference in the way the two codes. 
Starburst99 computes the isochrones from the implemented evolutionary
tracks and obtains the results from the implemented atmosphere library. 
In contrast, Sed@ currently does not compute isochrones, but takes them 
from the literature if they are in the appropriate format. 

\subsection{Stellar initial mass function (IMF) and star formation law}

The results presented here assume a power law IMF,
$\Phi\rm(m)=C m^{\alpha}$, where the constant $\rm C$ is determined by the
total gas mass converted into stars. The slope ${\alpha}=-2.35$ is
that derived by \cite{Sal55}. The low-mass and high-mass cutoffs have been chosen as
$M_{\rm low}=0.1\, \ms$ and $M_{\rm up}=120\, \ms$, respectively
\cite[see][for a detailed discussion about the more appropriate choice of the
lower mass limit]{GB93}. The IMF is binned into
a grid of the initial masses determined by the particular isochrone, and all the
stars belonging to the same mass bin are assumed to have exactly the same
properties. The synthesis is done directly by adding the different contributions from
the stars to all the synthesised wavelengths. Here, we explore only
instantaneous burst models, normalized to $1 \ms$.

\subsection{Stellar evolution and isochrone integration}

The models have been computed using the isochrone technique to produce
a smooth spectral evolution. This technique was introduced by \cite{CB91}
to avoid abrupt changes of the properties of the stellar populations
associated with changes in the evolutionary stages of the stars when the mass
resolution of the stellar tracks is inadequate. The isochrone synthesis
technique is different from the one used by \cite{TG76}, takes into
account the lifetime of the stars in each evolutionary
phases considered, imposing that the number of stars in a given phase of the stellar
evolution must be proportional to the life time of such phase of an
individual star. Such a condition, the so called {\it Fuel Consumption
Theorem, FCT} \cite[][]{RB86,Buz89}, is specially relevant for the faster
phases of the stellar evolution. Although both methods should be
equivalent, the results depend on the method used  
\cite[see][for an extensive discussion]{Buz95,Buz02b}.

In this work, two different sets of stellar isochrones are used.  The
first one is the set of isochrones
from the Padova group by \cite{Beretal94} and \cite{Gietal00} presented in
\cite{Gi02}\footnote{The full set is available at {\tt http://pleiadi.pd.astro.it/}}, with
metallicities Z=0.019 (solar), 0.008 and 0.004 (hereafter Padova
isochrones). The initial mass resolution of the used Padova isochrones is
sufficient to assure a correct integration of the isochrone-IMF product.  These
models follow the evolution from the zero-age main sequence to the early
asymptotic giant branch (AGB) and to the first thermal pulse, or to the
carbon ignition depending on the initial mass. They cover an age range from 4 Myr to 17 Ga
in a set of 74 isochrones with a 0.06 log time sampling. An
additional set at Z= 0.030 and 0.002 is also available on the Padova web server
but it comprises a smaller range of ages (0.063 to 15.8 Gyr). We note
that there are different sets of isochrones released by the Padova group, in particular, 
the isochrones from \cite{MG01} fulfill the FCT requirements, unfortunately such isochrones do not
cover the early ages of the evolution of stellar clusters. In this work we only present the results
for the quoted \cite{Beretal94} and \cite{Gietal00} set of isochrones.

Our second set has been computed from the list of non-rotating tracks of the
Geneva group 
\cite[][]{Schalleretal92,Schaetal93a,Schaetal93b,Charetal93}, including standardmass-loss
rates computed at five metallicities: Z=0.040, 0.020 (solar), 0.008, 0.004
and 0.001\footnote{Note that the default in Starburst99 are the "enhanced", 
not the "standard" mass-loss rate.}. These tracks include all the phases of the stellar evolution
from the zero-age main sequence to the end of the giant branch or the carbon
ignition.  The Geneva isochrones are calculated by means of a parabolic
interpolation between the tracks (log M, log t$_k$) in the HR diagram  on a variable
mass grid as prescribed in \cite{Cervetal01}. This prescription is also
used for the isochrone computations in Starburst99.  The isochrones
were generated at specific ages with a variable time step: $\Delta
t=1$ Myr for 1-10 Myr, $\Delta t=5$ Myr for 10-100 Myr, $\Delta t=100$ Myr for
100 Myr to 1 Gyr, and $\Delta t=1$ Gyr for 1-10 Gyr.  Unfortunately this 
computation of the isochrones did not assure the FCT requirements (but it
is partially implicit in the use of the parabolic interpolations), neither
did it 
include the evolution of stars in the horizontal branch or later phases,
which are quite relevant for the resulting spectrum at ages older than
1Gyr\footnote{The Geneva group also produces evolutionary tracks from the begining 
of the horizontal branch to the first thermal pulse in the AGB phase
\cite[][]{Charetal96}. However, these evolutionary phases have not been 
included due to the variations of the number of equivalent 
evolutionary points depending on the initial stellar mass.}. 
We refer to \cite{Lanetal02} and V\'azquez \& Leitherer (2004) for other details about the
completeness of the evolutionary phases and their relevance in the
isochrones.

Since we are working with SSPs, we do not need to treat
chemical evolution self-consistently in the evolutionary tracks. Therefore each star generation
has the same metallicity during the evolution. On the other hand, the chemical 
evolution in the atmosphere of the individual stars has not been considered in the 
isochrone-atmosphere model assignation.

Because the isochrone metallicities do not coincide totally with the 
stellar atmospheres metallicities, we compute the models combining the 
isochrones and atmospheres as follow: 
isochrones at Z= 0.040 and 0.030 with the twice solar atmospheres; 
isochrones at Z= 0.020 and 0.019 with the solar atmospheres; 
isochrones at Z= 0.008 and 0.004 with the half solar atmospheres; 
and isochrones at Z= 0.001 with the 1/10 solar atmospheres. 

\section{Results}

We are presenting the model predictions for the spectral evolution of
single stellar populations. First, we describe the spectral energy
distribution in the 3000--7000 \AA\ range. Then, we present the spectral
evolution of the hydrogen and helium lines (section 4.2) and of the metallic
lines (section 4.3). The effect of metallicity on the spectra is
illustrated, and the models are discussed for two different 
isochrone sets (Geneva and Padova).

\subsection{Spectral evolution}

Figure 2 shows the spectral evolution of a single stellar population with
solar metallicity at seven different ages of 4 Myr, 10 Myr, 50 Myr, 200 Myr, 1
Gyr, 3 Gyr, and 10 Gyr.  The first effect to notice is the decrease with time 
of the stellar population luminosity, due to the most
massive stars dying off. The second effect is the change in the shape of
the continuum with time, up to a few Gyr; this is a consequence of the large
variation of the effective temperature of massive and intermediate mass
stars during their evolution. However, after 3 Gyr, the shape of the
optical continuum is almost constant because low mass stars evolve within a
small range of effective temperature.  This results from the strong
concentration of the isochrones at ages older than 3-4 Gyr (see Figure 1). 
Additionally, Figure 2 illustrates the strong time evolution of
the Balmer and the 4000 \AA\ breaks. The evolution of the strength of the
Balmer lines is also striking. They have a maximum around 400 Myr, when 
A stars dominate the optical continuum.

\begin{figure*}

\hbox{
{\rotatebox{270}{\includegraphics[width=10cm]{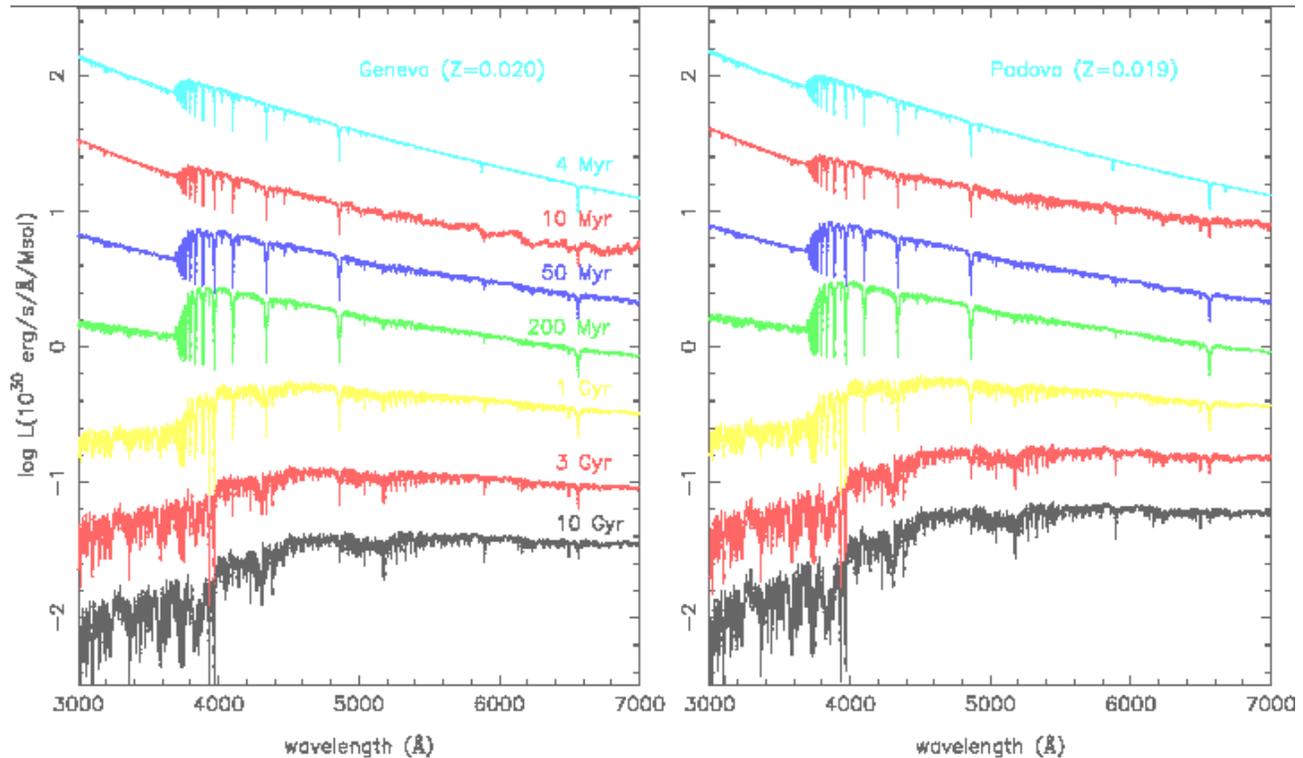}}}}
\caption{Spectral evolution for a single stellar population model at solar metallicity
following the Geneva (left) evolutionary tracks with standard mass loss, and the 
Padova (right) tracks. Ages from top to bottom are 4 Myr, 10 Myr, 50 Myr, 200 Myr, 
1 Gyr, 3 Gyr, and 10 Gyr.    }
\end{figure*}

Figure 2 also shows that Geneva and Padova models have similar
overall spectral characteristics; however, several detailed differences are
clearly visible in this figure. First, the spectra obtained with the Padova
models are always more luminous than the corresponding Geneva
spectra. Second, the continuum shapes diverge in several stages of the
evolution. For example, at older ages ($\geq2$ Gyr), the Padova spectra are
systematically redder than Geneva's due to the inclusion of more evolved
phases after the tip of the red giant branch, and the absence of low-mass 
stars\footnote{Note that the Geneva tracks are truncated at 0.8 M$\odot$ while 
the Padova isochrones include stars with mass M$\geq$ 0.2 M$\odot$. 
According with V\'azquez \& Leitherer (2004), the differences between Padova 
and Geneva models are mainly due to this truncation. }. 
These phases and low-mass stellar tracks are not included in the
Geneva isochrones. However, at younger ages ($\leq100$ Myr), there is no
systematic change of the shape of the continuum, with the Padova spectra bluer
or redder than Geneva depending on the given age. These differences result
directly from the isochrones (Figure 3). For example, at 50 Myr, the main
sequence turnoff in the Padova isochrones is bluer than in Geneva, and
the coolest stars in the Geneva tracks have lower $T_{\rm eff}$ than in the
Padova models; these differences make the Padova spectrum bluer than the
Geneva spectrum at 50 Myr.  A significant difference between both sets of models 
is at RSG ages, i.e., 10 Myr at solar metallicity. At this age, the Geneva spectrum is steeper 
and shows stronger metallic bands at red wavelengths ($\lambda \geq$ 5500 \AA) than
the Padova spectrum. These differences are produced by a larger 
contribution of cool and low gravity stars (Teff $\sim$4000 K, and log $g \sim$ 0) 
to the total luminosity in the Geneva with respect to the Padova models (Figure 3).  

\begin{figure*}
\hbox{
{\rotatebox{270}{\includegraphics[width=12cm]{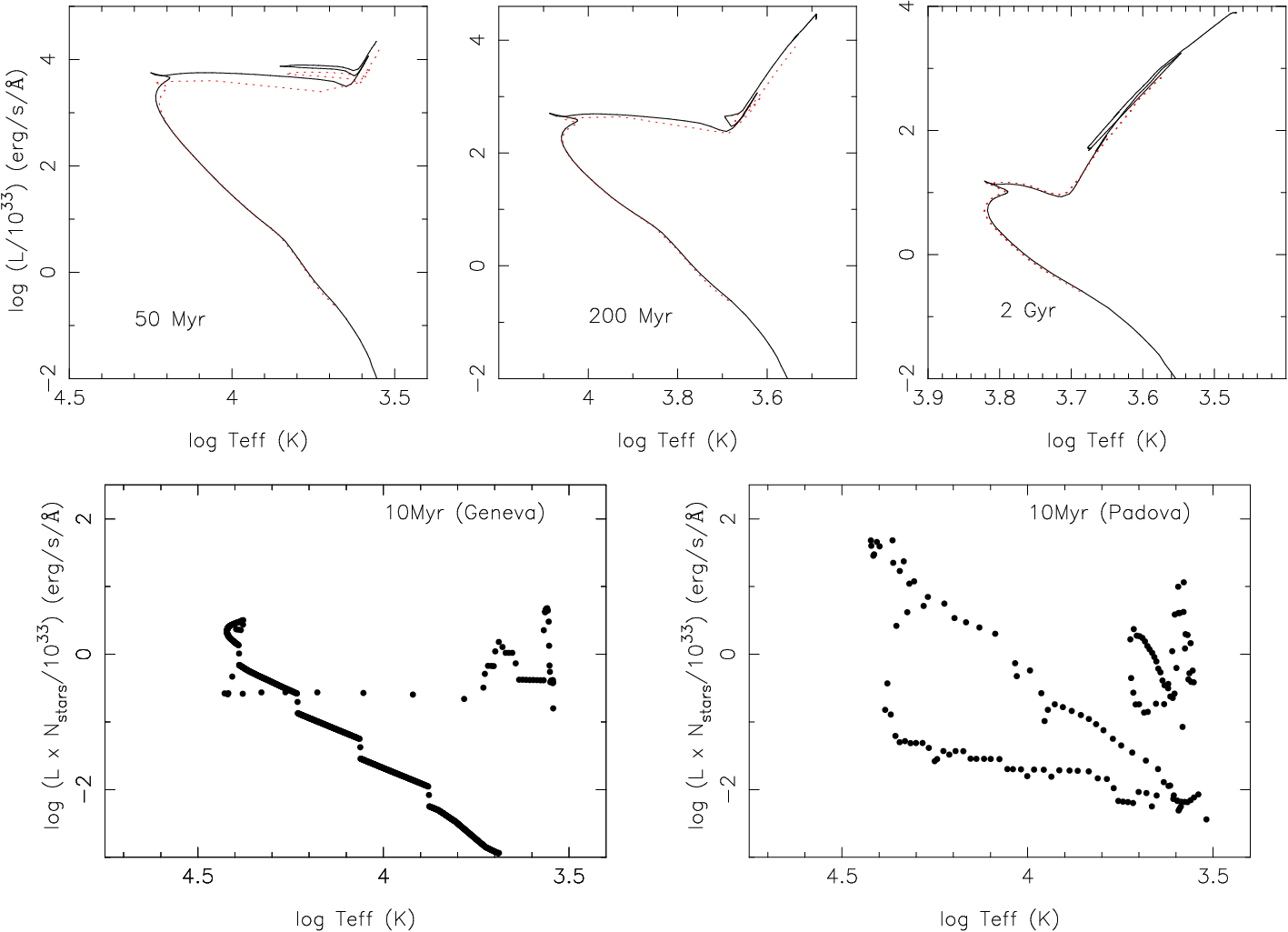}}}
}
\caption{Upper panel: Comparison of the isochrones obtained with the Padova (black line)
and Geneva (red dotted line) tracks for solar metallicity and ages of 50 Myr, 
200 Myr and 2 Gyr. Lower panel: Comparison of the Geneva (left) and Padova (right) isochrones 
at 10 Myr weighted by the number of stars in each luminosity bin. Note that the 
relative contribution to the total luminosity of stars 
with T$_{eff}$= 4000 K is significantly larger in the 
Geneva than in the Padova models. }
\end{figure*}

The metallicity effect on the spectral evolution is shown in Figure 4. We
plot the spectra of a 1 Gyr stellar population at four different
metallicities. The effect is appreciated in the slope of the continuum and
in the strength of the lines. At this age, lower metallicity 
spectra are bluer, have stronger Balmer lines, and weaker metallic lines (Figure
4a). These changes are produced by the shift of the isochrones to higher
temperatures with decreasing metallicity, which is a consequence
of stars being brighter and hotter at lower metallicity.  The effect of
metallicity is also evident in spectra at older ages ($>$few Gyr), where
the continuum shape is similar in all the models, but there is a
strengthening of the metallic lines from the most metal-poor to the most
metal-rich stellar populations (Figure 4b).

\begin{figure*}
\hbox{
%{\rotatebox{270}{\includegraphics[width=6.5cm]{rgdf4a.eps}}}
%{\rotatebox{270}{\includegraphics[width=6.5cm]{rgdf4b.eps}}}
{{\includegraphics[width=6.5cm]{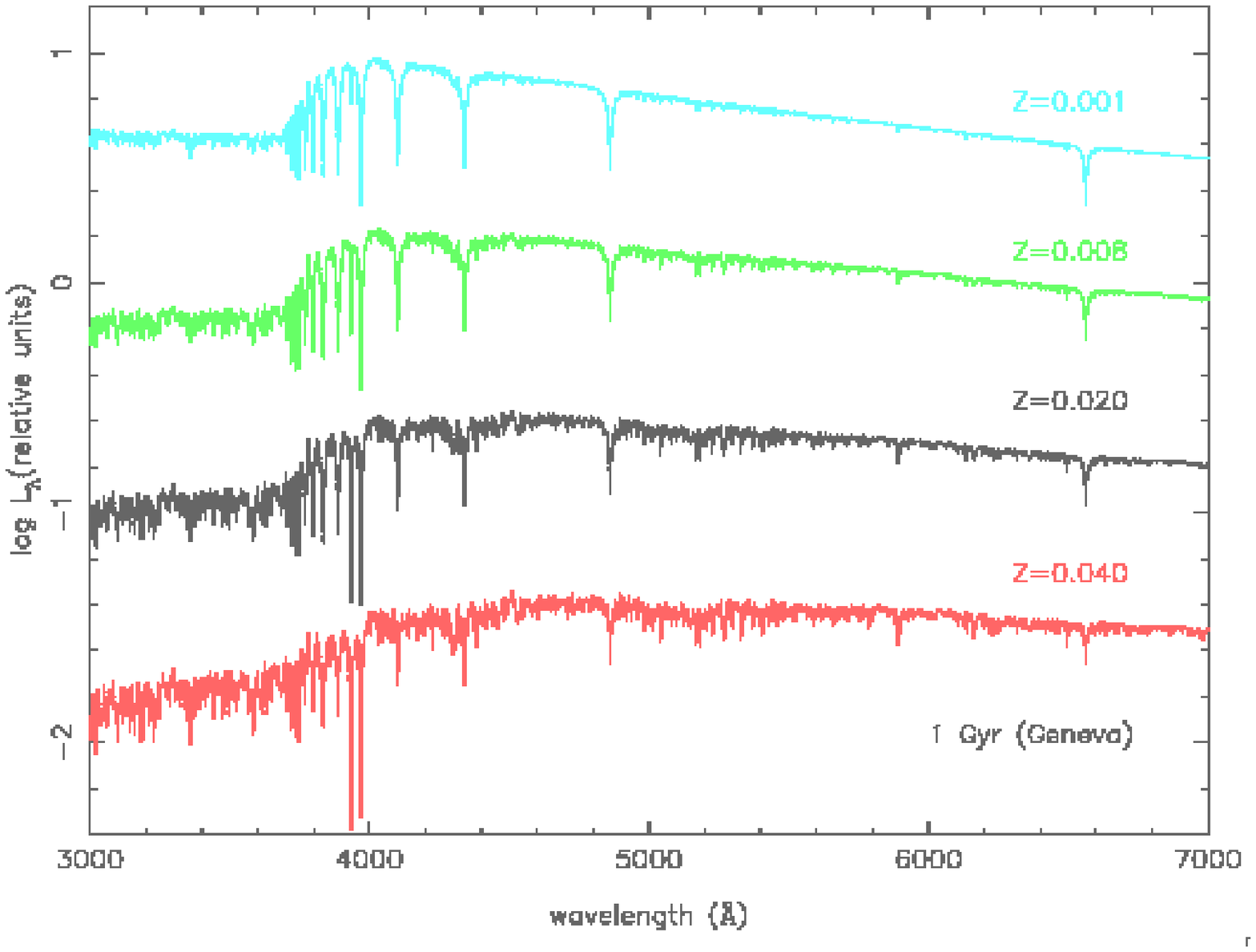}}}
{{\includegraphics[width=6.5cm]{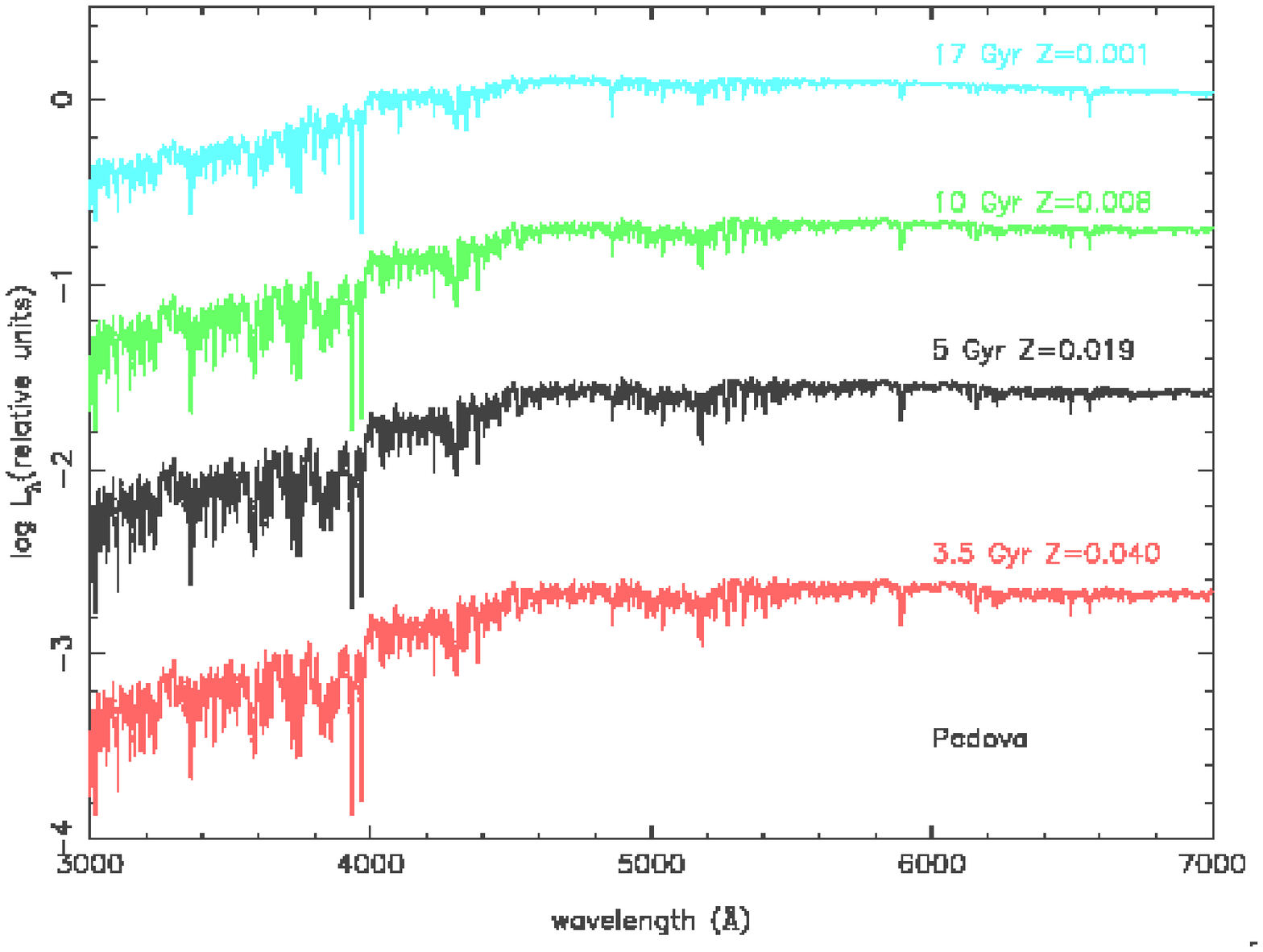}}}
}
\caption{Spectra predicted for a single stellar population at different ages for different metallicities 
(as labeled), obtained with the Geneva (left) and Padova (right) isochrones.}
\end{figure*}

\subsection{Helium lines}

The high spectral resolution of our stellar library allows the prediction of
the strength of many weak helium absorption lines that form in the optical continuum of
young stellar populations.  In Figure 5, we have labeled some of the most
relevant He{\sc i} and He{\sc ii} lines in the spectrum of a 7 Myr stellar
population at Z=0.001. They are: He{\sc i} $\lambda$3819, $\lambda$4026,
$\lambda$4387, $\lambda$4471, $\lambda$4922, and $\lambda$5876; and He{\sc
ii} $\lambda$4200, $\lambda$4541, $\lambda$4686 and $\lambda$5412. He{\sc
ii} lines are detected only during the first few years of the
evolution; their equivalent widths decreasing with time from values of
about 0.5 \AA.  He{\sc i} lines are more prominent and are detected during
the first 100 Myr of evolution, when the optical continuum is dominated by
B stars. Some of these He{\sc i} lines have been calculated previously by us
(GLH99).

We measured the equivalent widths of three of these lines (He{\sc i}
$\lambda$4026, $\lambda$4471 and $\lambda$5876) in windows of 9-14 \AA,
integrating the flux from the pseudo-continuum obtained by a first-order
polynomial fit to the continuum windows defined in Table 1.  Figure 6 shows
the equivalent widths of these lines for a stellar population at solar
metallicity. He{\sc i} $\lambda$4026 is one of the stronger lines, while
He{\sc i} $\lambda$5876 is much weaker.  The He{\sc i} lines present a
maximum value between 20 Myr and 50 Myr when the stellar population is
dominated by B stars.  He{\sc i} lines do not form after 100 Myr, the
lifetime of B stars. The equivalent width has also a maximum between 7-15
Myr due to the presence of post-main sequence stars with $T_{\rm
eff}\leq8000$ K (see Figure 7 in GLH99).

Note, however, that these predictions for hot stars are estimated using static 
stellar atmosphere models \cite[][]{LH03}. They 
do not account for wind effects, which are important in very massive and 
evolved O stars. The lines most affected by winds are H$\alpha$ 
and He{\sc ii} $\lambda$4686. The wind effects are less important in 
the Balmer and He{\sc i} lines in the blue spectral range, and at low metallicity 
\cite[][]{Watlas}.

The main effect of the metallicity on the He{\sc i} lines results from 
the effect of the metallicity on the temperature of the turnoff,
with metal-poor stellar populations having a higher turnoff temperature.  The
equivalent width of He{\sc i} lines is a function of $T_{\rm eff}$ and
gravity, with a maximum at 20000 K and ${\rm log}\,g=5.0$ (GL99). Thus, the
variation of EW(He{\sc i}) with metallicity depends on the variation of the
turnoff temperature of the stellar population with metallicity.

\begin{table*}
 \centering
 \begin{minipage}{140mm}
  \caption{Hydrogen and helium lines and continuum windows}
\begin{tabular}{@{}llll@{}}
  \hline
Line & Line windows & Continuum window (blue) & Continuum window (red) \\
\hline
 H$\delta$     & 4092--4112; 4087--4117; 4070--4130 & 4012--4020 & 4158--4169 \\
 H$\gamma$     & 4330--4350; 4325--4355; 4310--4370 & 4262--4270 & 4445--4453 \\
 H$\beta$      & 4852--4872; 4847--4877; 4832--4892 & 4770--4782 & 4942--4954 \\
 H$\alpha$     & 6553--6573; 6548--6578             & 6506--6514 & 6612-6620 \\
He{\sc i} $\lambda$4026  & 4020--4031       & 4012--4020  & 4158--4169 \\
He{\sc i} $\lambda$4471  & 4464--4478       & 4665--4675  & 4464--4478 \\
He{\sc i} $\lambda$5876  & 5871--5880       & 5835--5845  & 5904--5912 \\
\hline
\end{tabular}
\end{minipage}
\end{table*}

\begin{figure}
\hbox{
{\rotatebox{270}{\includegraphics[width=5cm]{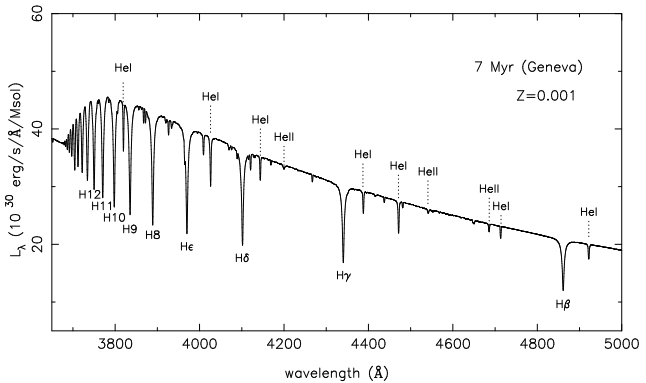}}}}
\caption{Spectrum predicted for a 7 Myr instantaneous burst at Z=0.001. 
Hydrogen and helium lines in the 3650--5000 \AA\ interval are indicated.}
\end{figure}

\begin{figure*}
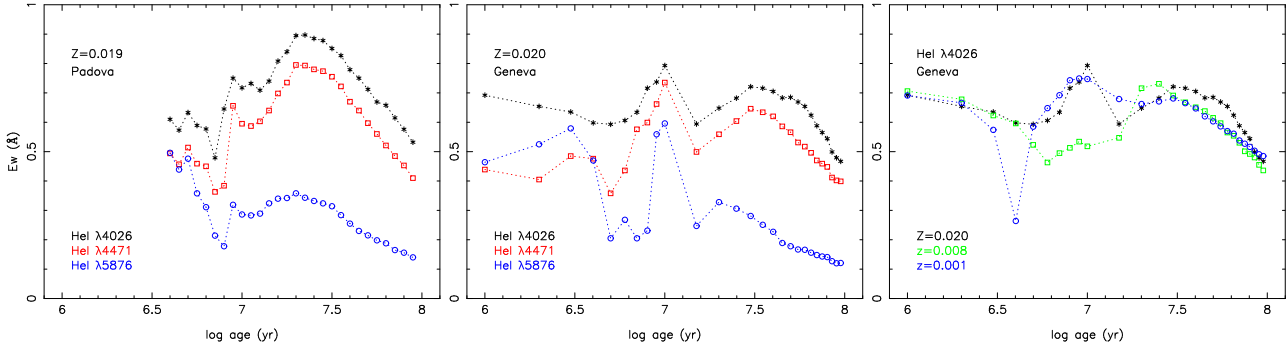

\hbox{
{\rotatebox{270}{\includegraphics[width=4.5cm]{rgdf6a.eps}}}
{\rotatebox{270}{\includegraphics[width=4.5cm]{rgdf6b.eps}}}
{\rotatebox{270}{\includegraphics[width=4.5cm]{rgdf6c.eps}}}
}
\caption{Equivalent widths of the He{\sc i} lines, $\lambda$4026 (stars, black), 
$\lambda$4471 (squares, red), $\lambda$5876 (circles, blue),  
for a SSP at solar metallicity for the Padova (left panel) and Geneva (center panel) stellar models. Right panel: Equivalent 
widths of the He{\sc i} $\lambda$4026 (Geneva tracks) for three metallicities, 
Z= 0.020 (stars, black), 0.008 (squares, green) and 0.001 (circles, blue). }
\end{figure*}

\subsection{Hydrogen lines}

Figure 7 shows the synthetic profiles of the Balmer lines (H$\alpha$,
H$\beta$, H$\gamma$, H$\delta$, H$\epsilon$, H8, H9, H10,..., to the Balmer
limit) for stellar populations at three different ages, 10 Myr, 100 Myr and
500 Myr at half solar metallicity, and for 1 Gyr at several metallicities
(Z=0.019, 0.008 and 0.001).  As it is well known, the profiles of the
Balmer lines change significantly with age. The tendency is that the width
of the lines increases with the evolution until about 500 Myr if the stellar
population has solar metallicity. At ages younger than 1 Gyr, the profiles
of these lines have only a small dependence on the metallicity. In 
individual stars, the strength of the hydrogen lines is independent of
metal content. Therefore, the lines depend on metallicity only through the dependence of
the stellar evolution in the integrated light of a stellar population.  

Since at low metallicity a star of a given mass is
hotter and evolves more slowly in the HR diagram, the maximum strength of the
Balmer lines occurs at older ages. However, the dependence of the indexes
of the Balmer lines with metallicity is stronger (Figure 8) at ages older
than 1 Gyr.   So, a metal-poor old stellar population can have Balmer lines
similar to an intermediate age stellar population at solar or over-solar.

\begin{figure*}
\hbox{
{\rotatebox{270}{\includegraphics[width=10cm]{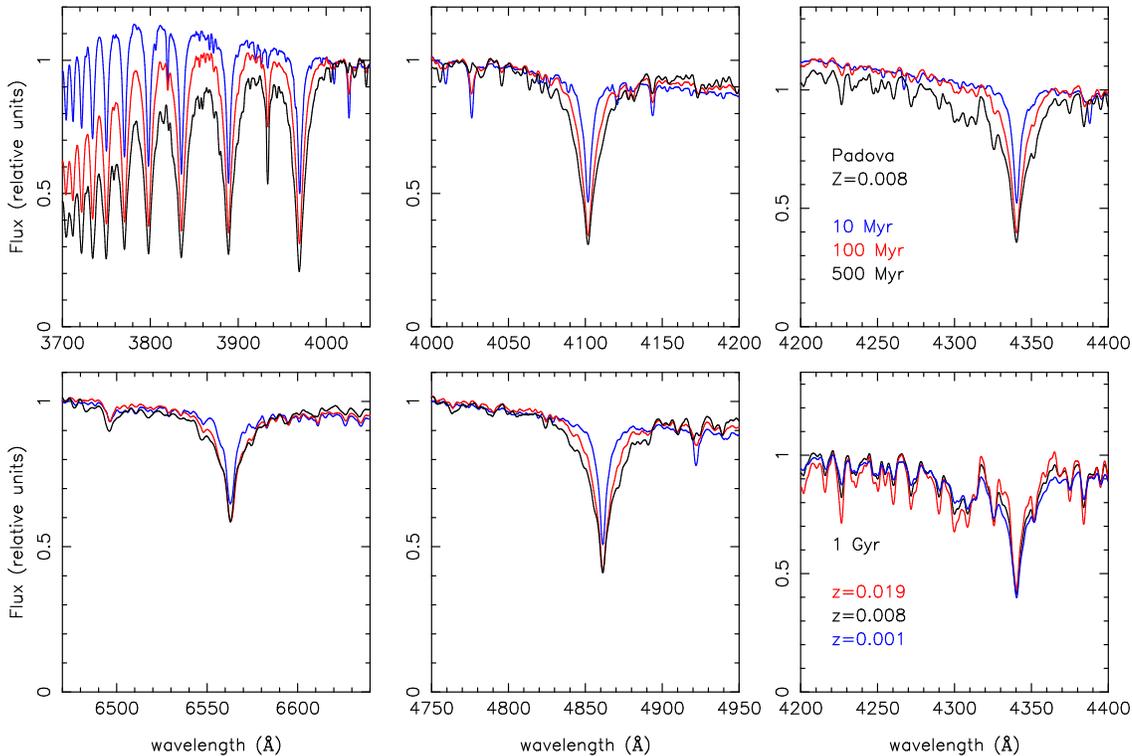}}}}
\caption{Synthetic spectra from 3700 to 6700 \AA\ predicted for SSPs at ages 
10 Myr (blue), 100 Myr (red), and 500 Myr (black), computed with the Padova isochrones at metallicity Z=0.008. 
Lower-right panel: H$\gamma$ at 1 Gyr and at three metallicities, Z=0.019 (red), 0.008 (black) and 0.001 (blue). }
\end{figure*}

Figure 8 shows the evolution of the equivalent widths of the Balmer lines
(H$\alpha$, H$\beta$, H$\gamma$ and H$\delta$). The equivalent widths of
these lines are measured in windows of width 20 \AA, 30 \AA, and 60 \AA,
centered on the lines, except for H$\alpha$ which is measured in only two windows.  
However, only the values corresponding
to the window of 30 \AA\ are plotted in Figure 8. These windows have been chosen to
evaluate the effect of line broadening on the EW. 
A narrow window is required to date old stellar populations, because it includes less
contribution of neighboring metallic lines; in contrast, a wide window that 
includes the entire line wings is more useful to date intermediate age populations.  
The measurements are obtained by integrating the flux pseudo-continuum
found through a first-order polynomial fit to the continuum windows defined
in Table 1. H$\beta$, H$\gamma$ and H$\delta$ have a very similar behavior
if the stellar population is younger than 1 Gyr. Their strengths range from
3 to 12 \AA. During the first few million years, the EW is almost constant
at $\sim3$ \AA.  Then, the equivalent width increases with time until 400 Myr,
when stars of 10000--9000 K dominate the optical continuum.  The time
evolution of H$\alpha$ is smoother, changing from 3 to 6 \AA.  Note that
the evolution of the strength of these lines is very similar in the Geneva
and Padova spectra if the stellar population is younger than 1 Gyr.  However,
Padova models will predict younger ages than the Geneva models in old
stellar populations.

We have not plotted the equivalent widths of the high-order Balmer
series. They have a similar behavior as H$\beta$ and H$\gamma$, with
values that range from 2-3 to 10 \AA\ (GLH99). These lines are very useful
for constraining the stellar ages in the nuclei of galaxies with an important
contribution from nebular emission \cite[e.g.][]{GHL01} due to the strong
decrease of the nebular emission with the Balmer decrement, while the EW of
absorption lines is constant or increases with wavelength.

\begin{figure*}
\hbox{
{\rotatebox{270}{\includegraphics[width=10cm]{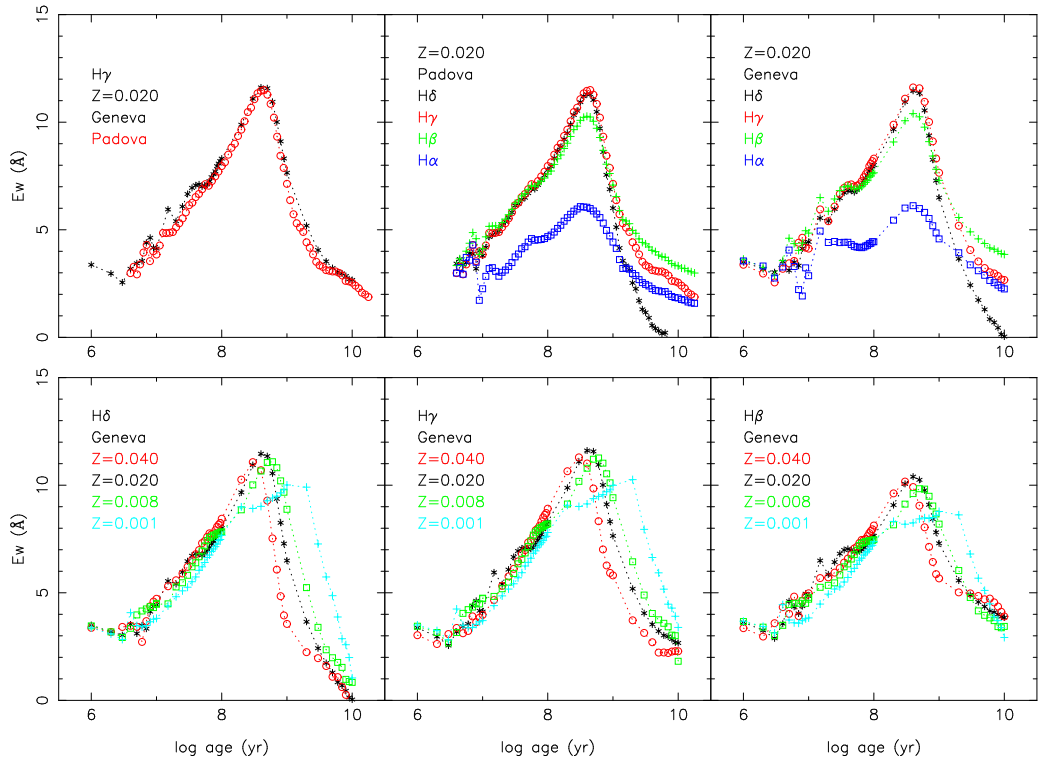}}}}
\caption{1) Lower panels: Equivalent widths of H$\delta$ (left), H$\gamma$ (central) and H$\beta$ (right)
at different metallicities, Z= 0.020 or 0.019 (stars), Z= 0.040 (circles), Z= 0.008 (squares), and Z= 0.001 (crosses)
from the Geneva tracks.
2) Upper left panel: Equivalent widths of H$\gamma$ at solar metallicity from the Geneva 
(stars) and Padova (circles).
3) Central and right upper panels: Equivalent widths of the Balmer lines, H$\alpha$ (squares), 
H$\beta$ (crosses), H$\gamma$ (circles) and H$\delta$ (stars), 
predicted from the Padova and Geneva tracks at solar metallicity.}
\end{figure*}

\subsection{The 4000 \AA\ break}

The Balmer discontinuity is another important age diagnostic in the optical
continuum of a stellar population.  Figure 2 shows how this discontinuity
evolves with time up to converging with the 4000 \AA\ break.  While the
4000 \AA\ break is produced by the accumulation of metallic lines around
4000--3800 \AA, the Balmer discontinuity depends on the accumulation of
the Balmer series below 4000 \AA. Different definitions can be adopted to
measure this discontinuity, and the usefulness of the definition depends on
the stellar age that we are interested in measuring. In section 4.5, we will
adopt the \cite{Rose94} definition for measuring the Balmer discontinuity to date
young stellar populations. However, we adopt the 4000 \AA\ break definition 
introduced by \cite{Balogh} to study the general change of the
discontinuity for intermediate and old stellar populations. Balogh et al. use 100
\AA\ continuum bands (3850--3950 \AA\ and 4000--4100 \AA) to measure
the break. The main advantage of this definition is that the index is less
sensitive to reddening effects. Other definitions using broader continuum
bands have been proposed by \cite{Bru83} and \cite{Getal99}.

Figure 9 shows the strong variation of the index for ages older than
100 Myr. This index displays a similar behavior for the Padova and
Geneva stellar models. However, the predictions differ for stellar
populations older than 1 Ga; Geneva models, which do not contain all
the relevant evolutionary phases, predict older ages, similarly to
the case of the Balmer lines. On the other hand, for ages older
than 400 Myr, this index definition significantly depends on the metallicity,
 mainly due to the fact that the blue continuum window
includes the CaII K line which starts to be prominent at intermediate
and old ages. The blue window also includes a CN band, which is strong
in metal-rich old stellar populations.

Recently, the 4000 \AA\ break has been used in combination with the H$\delta$ absorption line
to date stellar populations of galaxies \cite[][]{K03a,K03b,K03c}. In that work, the authors 
use the H$\delta_A$ 
definition by \cite{WO97}. This index uses a central (4082--4122 \AA) bandpass bracketed
by two pseudo-continuum band passes (4030--4082 \AA\ and 4122-4170 \AA). The evolution of the
4000 \AA\ break and the H$\delta_A$ index are plotted in Figure 9. Padova and Geneva models 
show similar values in the D$_{4000}$--H$\delta_A$ plane for ages younger than 2 Gyr, but, again, 
the Geneva models
predict older ages. This plot also shows the metallicity dependence of these two indexes. For a
stellar population younger than about 400 Myr, the D$_{4000}$--H$\delta_A$ plane does not 
depend on metallicity,
but for older ages, the two indexes change significantly as a consequence of the variation of the 4000 \AA\
break with the metal content and the effect of the metals on stellar evolution.

\begin{figure*}
\hbox{
{\rotatebox{270}{\includegraphics[width=10cm]{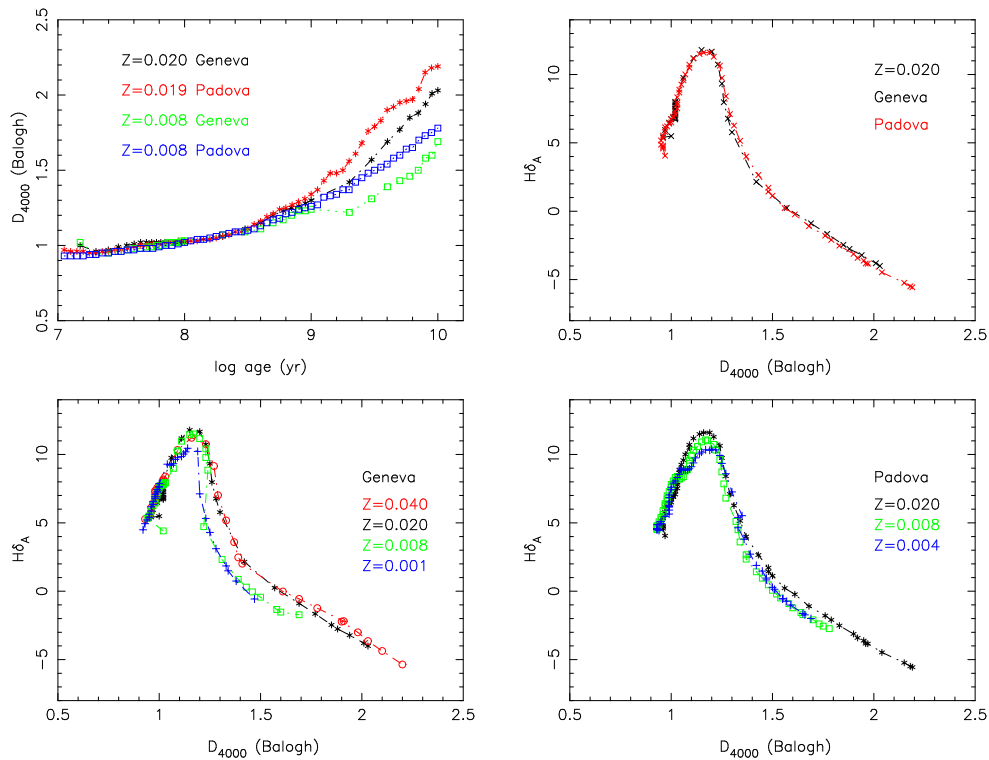}}}}
\caption{(1)  
Evolution of the 4000 \AA\ break (D$_{4000}$) measured in the synthetic spectra (upper-left panel) 
at solar (stars), and half solar (squares) metallicity for Geneva (black and green) and Padova (red and blue)
models. (2) Strength of H$\delta_A$ as a function of the 4000 \AA\ break for different
stellar models and different metallicities, Z= 0.020 or 0.019 (stars), Z= 0.040 (circles),
Z= 0.008 (squares), Z= 0.004 or Z= 0.001 (crosses), for ages between 10 Myr and 10 Gyr.  }
\end{figure*}

\subsection{Spectral indexes}

One of the most important problems in interpreting observed galaxy spectra 
is related to our ability to disentangle age and metallicity effects.
It is well known that the best way to decouple these two effects 
is by combining the strength of lines which are mainly sensitive to the main sequence turnoff temperature, 
like the Balmer lines, with  metallic lines which are sensitive to the temperature of the giant branch, which
depends on metallicity \cite[][]{W94,BMG94}.
The Lick system \cite[][]{Burstein84, Gorgas93,W94,WO97}, which includes more than 25 spectral 
indexes at optical wavelengths, was designed to break this age-metallicity degeneracy. These indexes 
are calibrated using Galactic stellar spectra observed at a resolution of $\sim9$ \AA, and
are useful to date stellar populations older than 1 Gyr. The lack of a significant number of 
high and intermediate mass stars in the stellar catalogs limits the ability of the system to
date young and intermediate age populations. However, owing to the high-spectral resolution of our new 
stellar library, the evolutionary models presented here are rather useful 
for investigating an optimum set of spectral indexes to disentangle the metallicity
and age effects in a large range of ages. This study will be presented elsewhere. 
Very recently \cite{Jietal04} have pointed out that the best 
way to break the degeneracy is by including all the spectral features blue-wards of 4500 \AA.

Alternatively, \cite{Rose84,Rose85} has proposed to use the ratio of the measure of the central line
depth of two neighboring spectral stellar features. This index definition has several advantages with 
respect to an equivalent width index. In particular, it is independent of the uncertainty associated with the 
location of the pseudo-continuum. In addition, this index is insensitive to reddening, but it depends
on the spectral resolution (Vazdekis 2001; Martins et al 2004). 
Many of these indexes have been used to constrain the stellar
population in elliptical galaxies \cite[e.g.][]{Rose94}, but also in young and intermediate age populations 
\cite[][]{LR03}. Two relevant indexes are H$\delta$/$\lambda4045$ and CaII. 
The former is defined as the ratio of the residual central intensity of H$\delta$ to that of
the FeI $\lambda$4045 line. The latter is defined as the ratio of the CaII H + H$\epsilon$ and CaII K lines.
These two indexes combined with the Balmer discontinuity, defined as the ratio of the average flux in two
narrow bands at 3700--3825 \AA\ and 3525--3600 \AA\ \cite[][]{RST87}, can constrain 
the age and metallicity of a stellar population. The behavior of these indexes with age and metallicity
is different. H$\delta$/$\lambda$4045 is very sensitive to age, and it shows a strong evolution from  
few 100 Myr to 10 Gyr (Figure 10), increasing with age. The minimum value occurs when the continuum 
is dominated by 10000 K stars, due to an enhancement of H$\delta$ in early A stars. Because stars of type
earlier than A0 have shallower Balmer lines, the index increases again toward younger ages. The CaII index
shows a similar behavior to the H$\delta$/$\lambda4045$ index at earlier ages, when the CaII
lines are weak and the ratio depends mainly on the H$\epsilon$ depth. Then, the index remains almost constant 
when the Balmer lines (age$\geq1$ Gyr) are weak and the core of the CaII lines are saturated 
(this happens in stars cooler than F5).
The CaII index is also more dependent on metallicity than the H$\delta$ and Balmer discontinuity. Figure 10
also shows the behavior of these indexes for two metallicities and for the Geneva and Padova models.

\begin{figure*}
\hbox{
{\rotatebox{270}{\includegraphics[width=10cm]{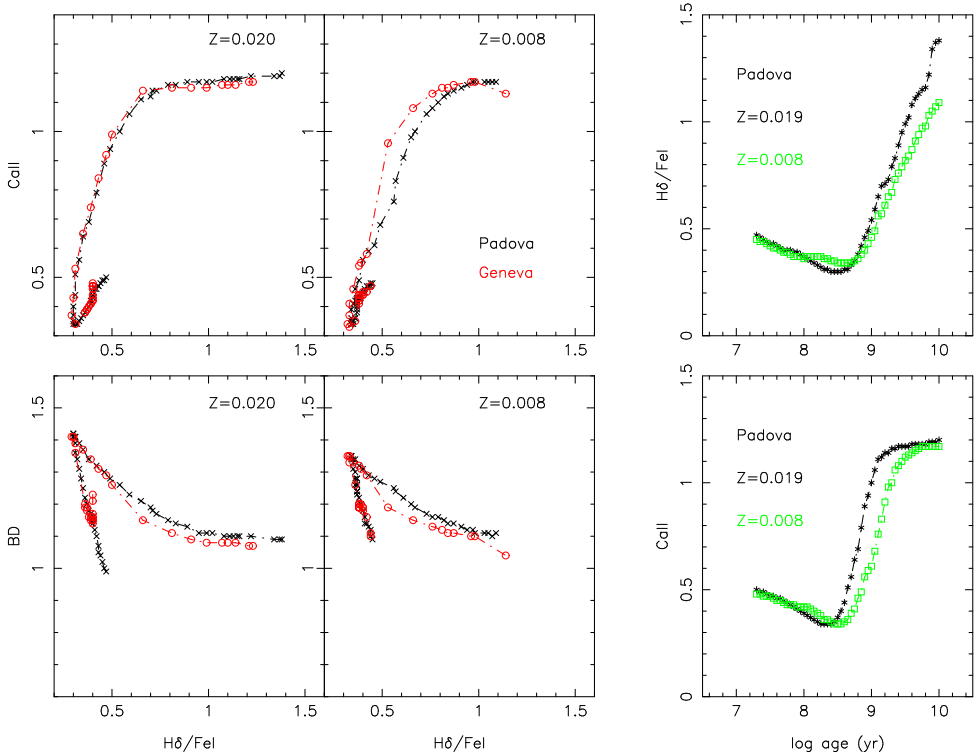}}}}
\caption{Strength of Rose's index (CaII, H$\delta$/FeI $\lambda$4045) and 
of the Balmer Discontinuity (BD) as a function of age and metallicity. The indexes 
are measured in the synthetic spectra that follow the Padova (black) and Geneva (red) models, 
at solar (left panel), at half solar (central panel) metallicities. In the right panels, 
the indexes plotted are from the Padova models, at Z= 0.019 (black, stars) and Z= 0.008 (green, squares).
Note that after 2-3 Gyr, the CaII index is constant and it shows a very small dependence with the metallicity
(see also Vazdekis 1999).}
\end{figure*}

\section{Comparison with previous work}

\subsection{Low spectral resolution models}

As pointed out in section 2, the new high spectral resolution stellar library  
presents two important changes with respect to previous work, related to the inclusion 
of line-blanketed models for hot ($T_{\rm eff}\geq27500$) and cool ($T_{\rm eff}\leq4500$) stars. 
To evaluate
these improvements and to check the changes in the shape of the continuum, we have compared the synthetic 
high resolution spectra with low-resolution models obtained by Starburst99  and 
{\sc galaxev}. Figures 11-13 show the comparison at three different ages, 1 Myr, 7 Myr and 5 Gyr. 

At 1 Myr (Figure 11), the optical continuum at $\lambda\geq$3750 \AA\ follows quite well the continuum, 
obtained with 
Starburst99 using Lejeune \cite[][]{Leje}, or Kurucz, stellar atmospheres. In the near-UV, however, the flux
is lower, and it is comparable to Starburst99 + UCL stellar atmospheres \cite[][]{smith}.
UCL models include O and W-R atmospheres with line blanketing by \cite{Pauldrach} and \cite{Hillier}. 
This difference between the fluxes at near-UV is clearly illustraded in Figure 13 of \cite{LH03}. 
They show that O stars with Teff$\geq$ 35000 K have lower emission in the 3000-4000 \AA\ 
interval than the Kurucz models. 
At older ages (not shown), the agreement between Sed@ (Geneva) and Starburst99+Lejeune is remarkable. 

Spectra corresponding to old stellar populations obtained with the Padova isochrones are also in very 
good agreement with the {\sc galaxev} low-resolution models which were computed using the \cite{Kur91}
stellar atmospheres models (Figure 12). 

However, there is a significant disagreement between the high resolution models
and low-resolution models at some specific ages when there is an important contribution of 
cool red supergiants, for example at 7-15 Myr in the Geneva, and at 9-14 Myr in the Padova models, 
both at solar metallicity (Figure 13). At these ages, the contribution to the total
luminosity of low gravity and cool (Teff$\leq$ 4000 K) stars (see the HR diagram in Figure 13) is 
very high. While the low-resolution models from Starburst99\footnote{Note, however, that this discrepancy
also appears if we compare these high-resolution models with Sed@ low-resolution models computed with 
the Geneva isochrones and Lejeune stellar atmospheres.}  
also use Kurucz and/or Leujene stellar
atmosphere models, here we are using PHOENIX models for these cool stars. Even though 
PHOENIX and Leujene atmospheres at low temperatures agree quite well for main sequence stars, 
PHOENIX produces less flux
than Kurucz at red wavelengths in stars of low gravity. This is the reason why the high-resolution models
have a steeper slope than the low-resolution models. See \cite{paperI} for further discussion.

\begin{figure}
\hbox{
{\rotatebox{270}{\includegraphics[width=5.5cm]{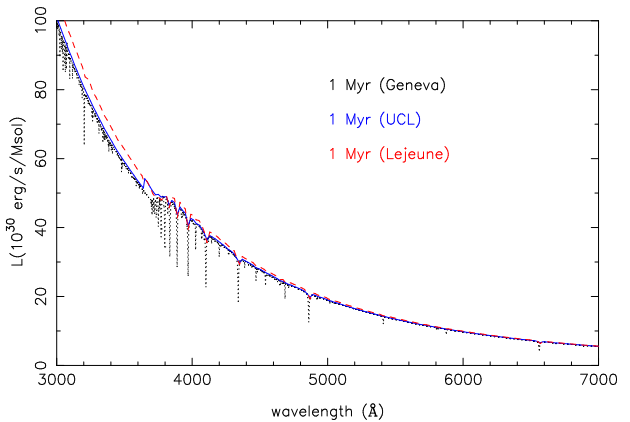}}}}
\caption{Comparison of the 1 Myr (Geneva) model at solar metallicity (black, dotted line) with the 
low-resolution Starburst99 models obtained with the stellar atmospheres of UCL (Smith et al. 2002) 
for massive and Wolf-Rayet stars (blue, continuum line) and Lejeune (red, dashed line).}
\end{figure}

\begin{figure}
\hbox{
{\rotatebox{270}{\includegraphics[width=5.5cm]{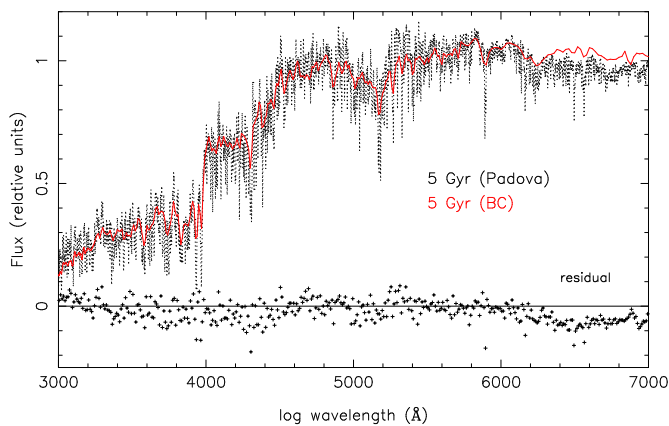}}}}
\caption{Comparison of the 5 Gyr (Padova) 
model at solar metallicity with the low resolution model from {\sc galaxev} (BC). 
The difference between the two models is shown as points around the line at y=0. }
\end{figure}

\begin{figure*}
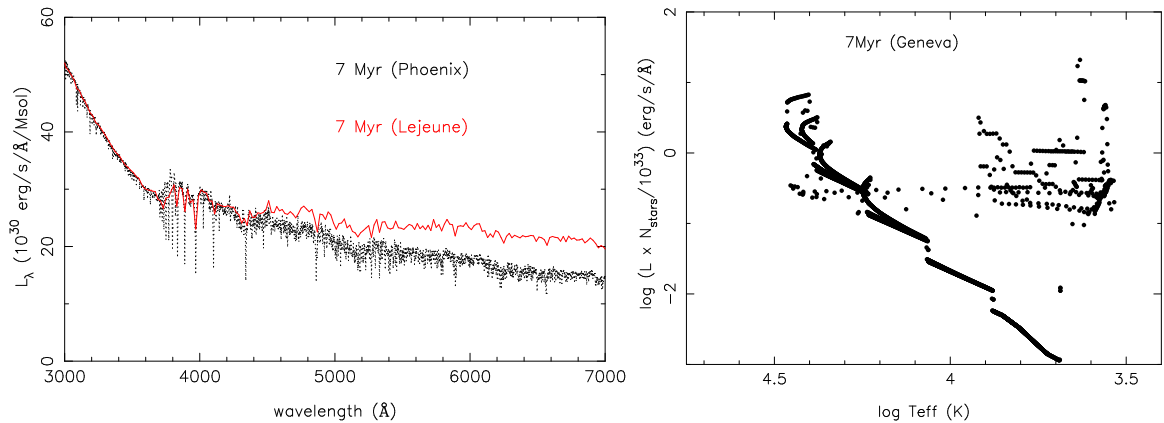

\hbox{
{\rotatebox{270}{\includegraphics[width=5.5cm]{rgdf13a.eps}}}
{\rotatebox{270}{\includegraphics[width=5.5cm]{rgdf13b.eps}}}
}
\caption{Comparison of the high (black-points) and low (red-continuum line) spectral resolution models 
for a 7 Myr instantaneous burst at solar metallicity (left). The models are computed 
with the Geneva tracks using Starburst99+Lejeune stellar atmospheres (red-continuum line)
and with Sed@ using our stellar library that includes PHOENIX atmospheres for Teff$\leq$ 4500 K (black-points).
Geneva isochrone at 7 Myr weighted by the number of stars in each luminosity bin (right). }
\end{figure*}

\subsection {Intermediate spectral resolution models}

Next, we compare our synthetic spectra with the available intermediate spectral resolution models 
by \cite{Vaz99} and {\sc galaxev}. These comparisons allow
a check in terms of the continuum shape but also in terms of the
stellar line strengths. 

\cite{Vaz99} computed single stellar population models at a resolution of $\sim1.8$ \AA\
in two reduced spectral regions 3850--4500 \AA\ and 4800--5450 \AA. He uses a subsample of 
$\sim500$ stars from the original \cite{Jones97} empirical stellar library. The models are 
computed at several metallicities and for ages older than a few 100 Myr.  In Figure 14, we compare
our and Vazdekis's models for a stellar population of 2 Gyr at solar metallicity in the 
two wavelength regions covered by \cite{Vaz99}. The agreement between the continuum shape 
and the strength of the lines in the two models is excellent.

\begin{figure*}
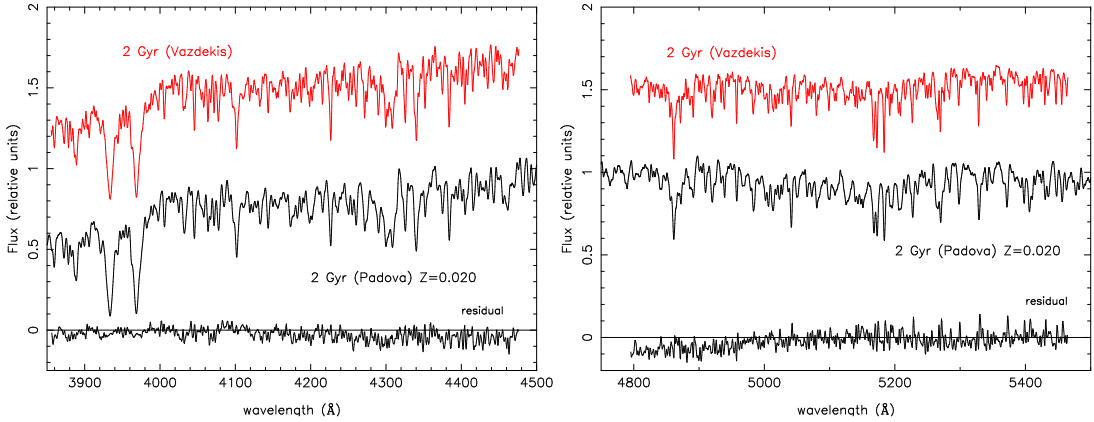

\hbox{
{\rotatebox{270}{\includegraphics[width=5.5cm]{rgdf14a.eps}}}
{\rotatebox{270}{\includegraphics[width=5.5cm]{rgdf14b.eps}}}
}
\caption{Comparison of the 2 Gyr (Padova) model at solar metallicity with
  the Vazdekis (1999) models.
The spectra are normalized at 4450 \AA\ (left) and at 5400 \AA\
(right). For the sake of clarity, the Vazdekis (1999) spectra are shown with a flux bias.
The residual spectrum is the difference between the two models.}
\end{figure*}

Recently, \cite{BC03}  presented 3 \AA\ resolution models covering from 3200 \AA\ to
9500 \AA\ ({\sc galaxev} code results),  a spectral range similar to our models. 
These predictions are based on the STELIB library 
\cite[]{LeBetal03}, which contains 249 stellar spectra at several metallicities 
($-2.0\leq{\rm[Fe/H]}\leq+0.5$), and spectral types from O to M9. However, the coverage in spectral
type is not uniform at all metallicities. In particular, there is a significant lack of massive stars
with 8000 $\leq T_{\rm eff}\leq$ 25000 and ${\rm log}\,g\leq$ 3.5. 
Figure 15 compares our and {\sc galaxev} models at solar metallicity for two ages, 11 Gyr and 25 Myr. 
There is excellent agreement between both models at old ages in terms of continuum shape 
and line strengths. However, at younger ages, both models differ in the line strengths.
The Balmer lines are stronger in our models, and the Balmer discontinuity is larger in our models.

\begin{figure*}
\hbox{
{\rotatebox{270}{\includegraphics[width=12cm]{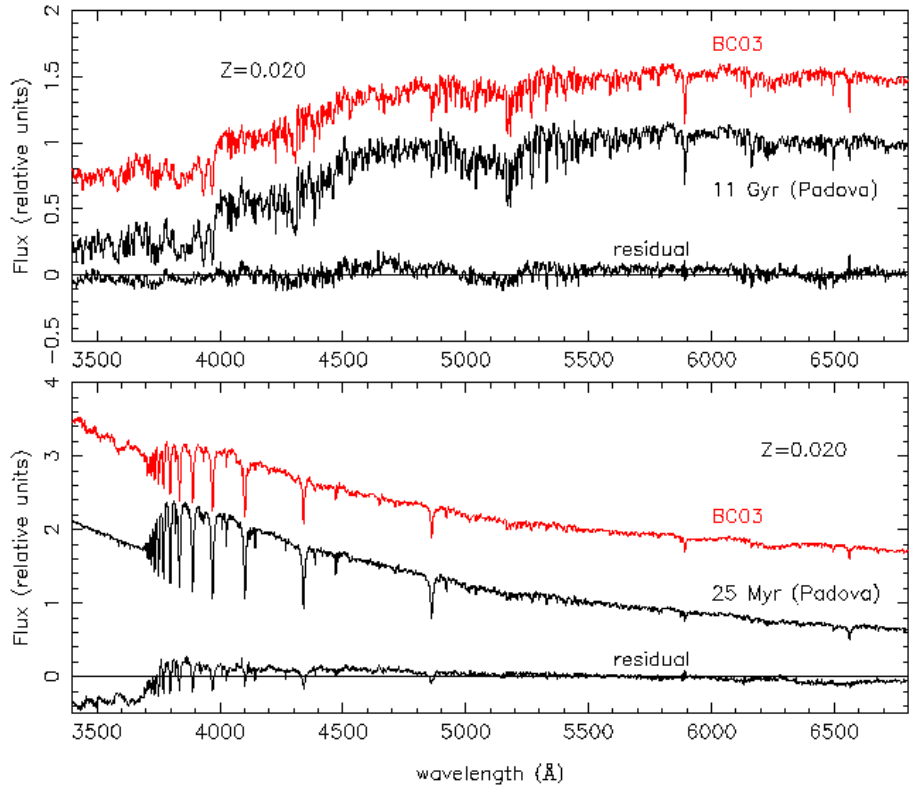}}}}
\caption{{\sc galaxev} models of a SSP of 25 Myr (bottom panel) and 11 Gyr (upper panel) 
at solar metallicity are compared with the high-resolution synthetic spectra presented here.
These models have been degraded to 3 \AA\ resolution, which is the resolution of the {\sc galaxev} models. 
The residual spectrum is the difference between the two models. }
\end{figure*}

\section{Comparison with LMC and SMC clusters}

In this section, we compare the continuum shape and strength of the stellar lines of our models 
with observations of stellar clusters in the Magellanic Clouds (LMC and SMC). \cite{LR03}
have obtained very good quality optical spectra of clusters in the LMC and SMC
covering a large range in metallicity (from [Fe/H]=+0.01 to -1.97) and ages (from 40 Myr to 15.5 Gyr).
The observations have an average spectral resolution of 3.2 \AA\ and cover the range  $\sim3420-4750$ \AA.
These data have been kindly provided by Dr. Rose. Figure 16 compares the spectra of three of these
clusters (NGC 1846, NGC 1831 and NGC 1818) with our models. Models and observations are flux normalized 
at 4010 \AA. The models are also downgraded to the spectral resolution of the observations and 
shown with a 0.5 flux bias for the sake of clarity. The age and metallicity reported 
for these clusters in the literature are labeled in the plot, as well as the values from our models. 
The agreement between observations and synthetic spectra is very good, considering an optimization of the
fit was not attempted.   

\begin{figure*}
\hbox{
{\rotatebox{270}{\includegraphics[width=12cm]{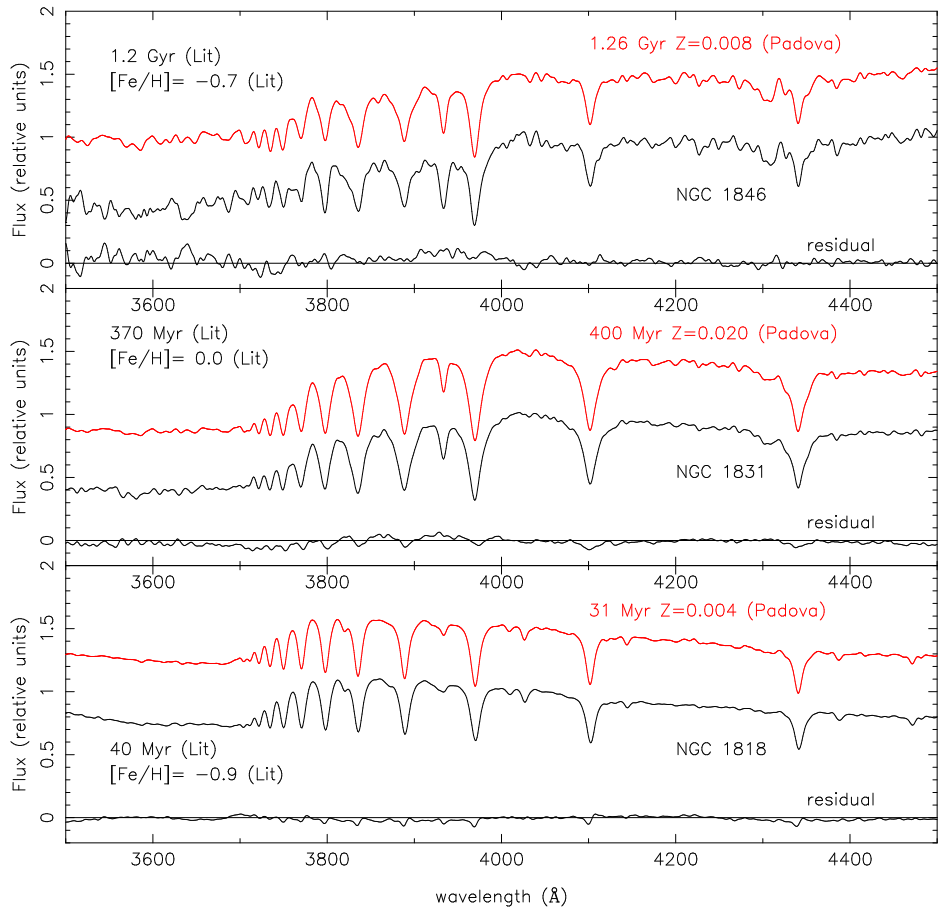}}}}
\caption{Intermediate resolution spectra of three LMC clusters,  
NGC 1818 (bottom), NGC 1831 (center) and NGC 1846 (top), are compared with the models 
(smoothed to the resolution of the observations). The metallicity 
and age reported in the literature and in the models are indicated. 
The spectra are flux normalized at 4010 \AA. LMC spectra 
are shown with a 0.5 flux bias for the sake of clarity. 
The difference between the data and the model fitted is shown as a residual line around y=0. 
An optimization of the fit was not attempted. }
\end{figure*}

\section{Limitations of the models in the Red Supergiant Phase}

In this section we discuss some of the limitations of the models that are directly 
related with the limitations imposed by the stellar evolution models.  
One of the larger uncertainties in the stellar models is in  
the red supergiant (RSG) phase, due to the uncertainties that exist associated with the 
mass-loss rates and mixing (Yi 2003). 
These limitations have been previously noted by the poor match of the NIR colors predicted by
starburst models with the integrated light of the observed stellar populations \cite[][]{Orietal99}.
More recently, \cite{MO03} have found that the Padova and Geneva models
are not able to produce RSGs as cool and luminous as what is actually observed. 
The stellar tracks of M$\geq$ 20 M$\odot$ produce RSGs of 6000-5000 K, 
but no RSGs stars in the LMC and SMC with T$_{eff}\geq$ 4500 K are observed. 

Even though at blue wavelengths RSGs may contribute less to the integrated light 
of a stellar population than in the red range, we have found a 
very important effect at some especific ages, e.g. 5 Myr (Geneva models)
 and 8 Myr (Padova) at solar metallicity. There is a strong broad feature around the
CaIIK line (at 3933 \AA) that is produced when RSGs of T$_{eff}$= 6000-5000 K contribute 
significantly to the continuum (Figure 17). This peculiar feature is another clear indication
that evolutionary tracks produce RSGs that are much hotter than those observed.
  
\begin{figure*}
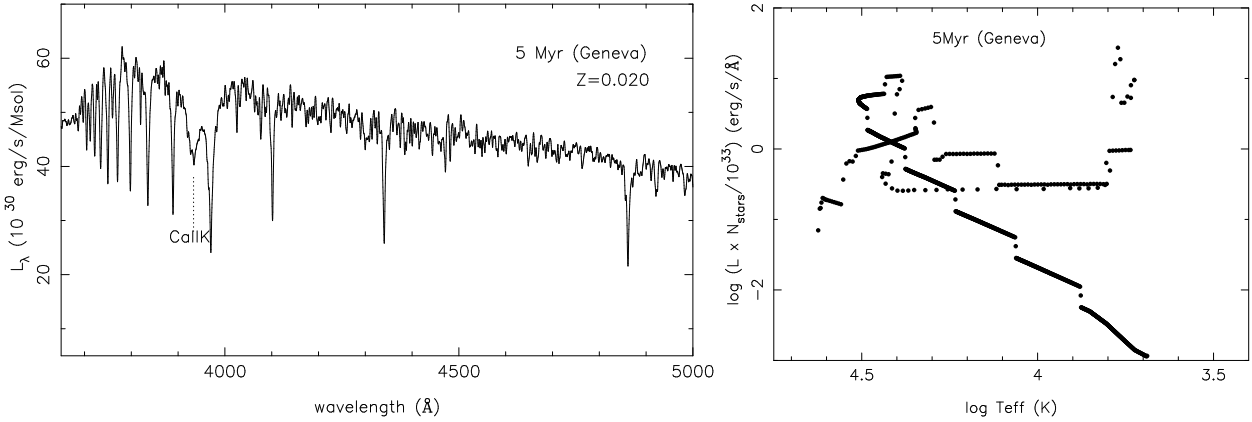

\hbox{
{\rotatebox{270}{\includegraphics[width=5.5cm]{rgdf17a.eps}}}
{\rotatebox{270}{\includegraphics[width=5.5cm]{rgdf17b.eps}}}
}
\caption{Spectrum (left) and isochrone (right) of a 5 Myr stellar population obtained with the Geneva 
isochrones at solar metallicity.  }
\end{figure*}

At older ages ($\geq$10 Myr), high-metallicity stellar models are able to produce RSG stars 
with T$_{eff}\leq$ 4000 K. However, low-metallicity stellar models also fail to reproduce 
the NIR colors observed in blue compact dwarf galaxies \cite[e.g.][]{VL04} and
stellar clusters in LMC and SMC that are in the RSG phase \cite[e.g.][]{Mayya97,Orietal99}. 
These results suggest clearly that isochrones are not adequate at low metallicity 
and at ages between 10-30 Myr.
For this reason, we have computed a set of evolutionary models that follow the stellar tracks 
at solar metallicity but where the stellar atmospheres have 1/10 and 1/2 solar metallicity. 
Even though this exercise can not solve the problem inherent in the stellar tracks, it can provide a
set of models that predicts some stellar characteristics that are more similar to those observed 
in stellar populations at low metallicity in the RSG phase. Figure 18
compares the spectra obtained with the Padova models that follow the isochrones at solar metallicity 
with the atmospheres at solar, half and 1/10 solar metallicity, and models obtained with the Z=0.004
stellar tracks. The main effect is seen in the slope of the spectra, almost identical in 
all the models that follow the solar tracks, but flatter, therefore redder, than the spectrum
that follows the tracks at Z=0.004. The variation of the equivalent widths of the metallic 
stellar lines, such as CaIIK and G band, is small at the RSG phase, and it is only important 
in stellar populations older than 1 Gyr (Figure 18). 

\begin{figure*}
\hbox{
{\rotatebox{270}{\includegraphics[width=10cm]{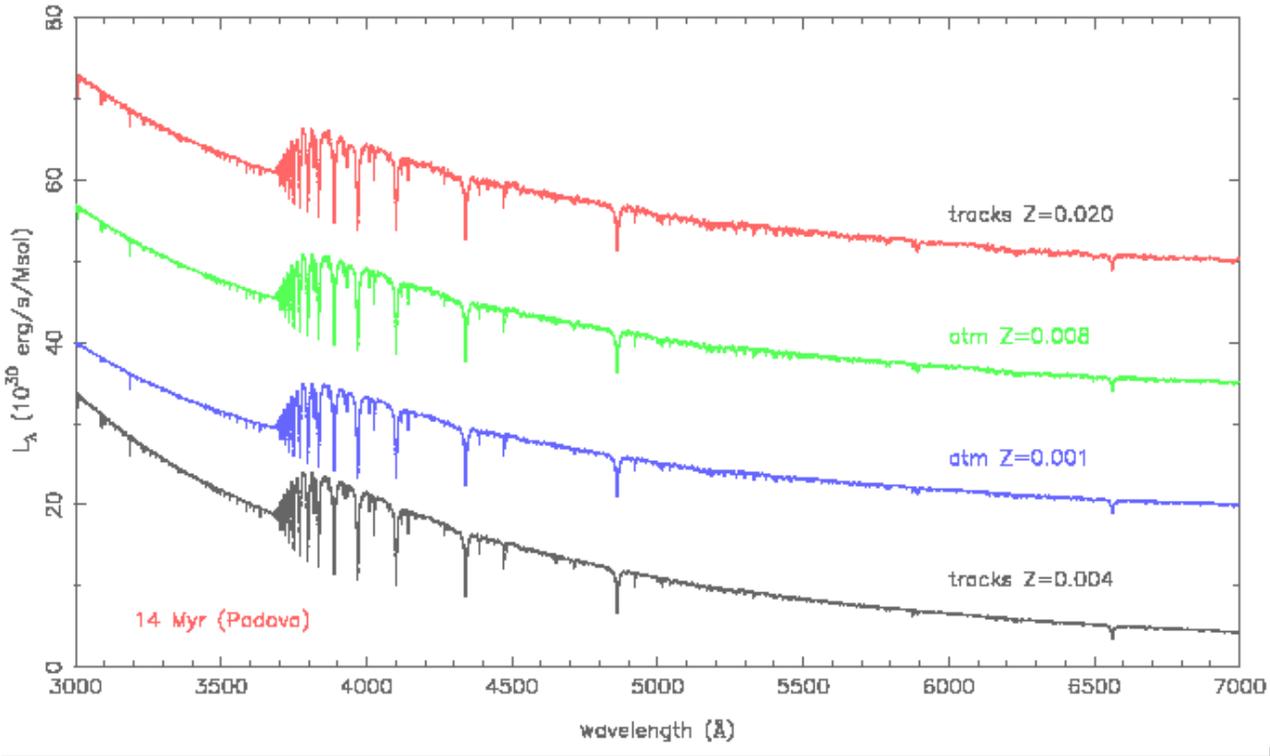}}}
}
\caption{Upper panel: spectra obtained with the Padova isochrones at several metallicities: 
isochrone at Z= 0.020 and the stellar atmospheres at solar (black), 
half solar (green) and 1/10 solar (blue) metallicity;
isochrone at Z= 0.004 and the stellar atmospheres with half solar metallicity (magenta).
These spectra are for SSP of 14 Myr. 
Lower panel: equivalent widths of the CaIIK (3933 \AA) line  
and the G band (4300 \AA). They are measured using the line and continuum windows
defined by Bica \& Alloin (1986). The colors represent models obtained with the isochrone and 
atmospheres with the metallicities indicated above.}
\end{figure*}

\section{Summary and Conclusions}

We have computed evolutionary synthesis models that predict high-resolution 
spectra of a stellar population in the 3000-7000 \AA\ range. These models 
predict the indexes of metallic lines and the absorption-line profiles of the
hydrogen Balmer series and helium lines as a function of age (1 Myr to 17 Gyr)
and metallicity (Z= 0.040, 0.020, 0.008, 0.004 and 0.001) for an 
instantaneous burst. The models are generated using the isochrone synthesis method.
These isochrones were built based on the Geneva and Padova evolutionary stellar tracks.
A new stellar library based on stellar atmospheres models is used to compute the 
near-UV and optical continuum. The library, composed of 1650 spectra, incorporates 
the lastest improvements in stellar atmospheres: non-LTE line-blanketed models for
hot stars (T$_{eff}\geq$ 27500 K) and PHOENIX LTE line-blanketed models for cool stars 
(3000 $\leq$T$_{eff}\leq$ 4500 K). Together with Kurucz models (for stars with
4750 $\leq$T$_{eff}\leq$ 27000 K), these atmospheres are used to generate high-spectral 
resolution spectra with a sampling of 0.3 \AA. 
The full set of evolutionary models and the stellar library spectra 
are available for retrieval at the websites {\tt http://www.iaa.csic.es/$\sim$rosa} and
{\tt http://www.iaa.csic.es/$\sim$mcs/Sed@}, or on request from the authors.

The results of the synthetic models with the Padova and Geneva isochrones are compared. 
The models obtain optimal results, or at least results as good as those of {\sc galaxev} or
\cite{Vaz99} for old ages. At such ages, the Padova isochrones work better but only
in the sense that they are more complete than Geneva in evolutionary phases.
They are better than previous work for young ages due to the larger
atmospheres coverage. Geneva and Padova produce quite similar results.
In general, age determinations are dependent on the isochrone assumptions.
These models have limitations mainly associated with the limitations and uncertainties
in stellar evolution. These limitations can be relevant in the RSG phase,
and in particular at low-metallicity.

Although the coverage is optimal, it is not correct to use {\it any} model for
clusters with total masses below ten times the Lowest Luminosity Limit established 
by \cite{CL04}; see also \cite{CVG03,Cetal03}.
This limit corresponds to the situation where the integrated light of the cluster equals 
to the luminosity of a single star in the model, and corresponds typically to cluster masses 
around $10^5$ M$_\odot$, but varies with wavelength and age \cite[see][for more details]{CL04}. 
The monochromatic Lowest Luminosity Limit and the intrinsic dispersion in the monochromatic 
flux due to incomplete sampling 
($\cal{N}$ or $N_{eff}$; see \cite{Cetal02} for details) are included in the output
of Sed@ and in the tables in the web server.

\section*{Acknowledgments}

We thank James Rose who kindly sent us the spectra of stellar clusters in the Magellanic Clouds, and 
Enrique P\'erez and Alejandro Vazdekis for very useful comments that improved the presentation of the paper. 
RGD, and MC acknowledge support by the Spanish Ministry of Science and 
Technology (MCyT) through grant AYA-2001-3939-C03-01 and AYA-2001-2147-C02-01.
PHH was supported in part by the P\^ole Scientifique de Mod\'elisation
Num\'erique at ENS-Lyon. Some of the calculations presented here were
performed at the H\"ochstleistungs Rechenzentrum Nord (HLRN), at the
National Energy Research Supercomputer Center (NERSC), supported by the
U.S. DOE, and at the San Diego Supercomputer Center (SDSC), supported
by the NSF. We thank all these institutions for a generous allocation
of computer time.

\bsp

\label{lastpage}


\begin{thebibliography}{}

\bibitem[\protect\citeauthoryear{Allard et al.}{2001}]{AHA01} Allard, F., Hauschildt, P.H., Alexander, D.R.,
Tamanai, A., \& Schweitzer, A. 2001, ApJ, 556, 357

\bibitem[\protect\citeauthoryear{Aretxaga et al.}{2001}]{Are} Aretxaga, I., Terlevich, E., Terlevich, R., 
Cotter, G., \& D\'\i az, A.I. 2001, MNRAS, 325, 636

\bibitem[\protect\citeauthoryear{Balogh et al.}{1999}]{Balogh} Balogh, M.L., Morris, S.L., Yee, H.K.C., Carlberg, R.G.,
\& Ellingson, E. 1999, ApJ, 527, 54

\bibitem[\protect\citeauthoryear{Bertelli et al.}{1994}]{Beretal94} Bertelli, G., Bressan, A., Bertelli, G., Chiosi, C., 
Fagotto, F., \& Nasi, E. 1994, A\&AS, 106, 275 

\bibitem[\protect\citeauthoryear{Bica \& Alloin}{1986}]{BA86} Bica, E., \& Alloin, D. 1986, A\&A, 166, 83

\bibitem[\protect\citeauthoryear{Brinchmann et al.}{2004}]{Brietal03} Brinchmann, J., Charlot, S., White, S.D.M., 
Tremonti, C., Kauffmann, G., Heckman, T., \& Brinkmann, J. 2004, MNRAS, 351, 1151

\bibitem[\protect\citeauthoryear{Bruzual}{1983}]{Bru83}
Bruzual, A.G. 1983, ApJ, 273, 105 

\bibitem[\protect\citeauthoryear{Bruzual}{2002}]{Bru00}
Bruzual, A.G. 2002, IAUS, 207, 616 

\bibitem[\protect\citeauthoryear{Bruzual \& Charlot}{2003}]{BC03} Bruzual, G., \& Charlot, S. 2003,
  MNRAS, 344, 1000

\bibitem[\protect\citeauthoryear{Burstein et al.}{1984}]{Burstein84} Brurstein, D., Faber, S.M., 
Gaskell, C.M., Krumm, N. 1984, ApJ, 287, 586

\bibitem[\protect\citeauthoryear{Buzzoni}{1989}]{Buz89}
Buzzoni, A. 1989, ApJS, 71, 817

\bibitem[\protect\citeauthoryear{Buzzoni}{1995}]{Buz95}
Buzzoni, A. 1995, ApJS, 98, 69

\bibitem[\protect\citeauthoryear{Buzzoni}{2002a}]{Buz02a}
Buzzoni, A. 2002a, AJ, 123, 1188 

\bibitem[\protect\citeauthoryear{Buzzoni}{2002b}]{Buz02b}
Buzzoni, A. 2002b, in New Quests of Stellar
Astrophysics, eds. M. Chavez et al., (Kluwer: Dordrecht) 189

\bibitem[\protect\citeauthoryear{Buzzoni, Mantegazza, \& Gariboldi}{1994}]{BMG94} 
Buzzoni, A., Mantegazza, L., \&  Gariboldi, G. 1994, AJ, 107, 513

\bibitem[\protect\citeauthoryear{Cervi{\~ n}o et 
al.}{2001}]{Cervetal01} Cervi{\~ n}o, M., G{\' o}mez-Flechoso, 
M.A., Castander, F.J., Schaerer, D., Moll{\' a}, M., Kn{\" o}dlseder, J., 
\& Luridiana, V. 2001, A\&A, 376, 422 

\bibitem[\protect\citeauthoryear{Cervi{\~ n}o \& 
Luridiana}{2004}]{CL04} Cervi{\~ n}o, M., \& Luridiana, V. 2004, 
A\&A, 413, 145 

\bibitem[\protect\citeauthoryear{Cervi{\~ n}o, Luridiana, \& 
Castander}{2000}]{CLC00} Cervi{\~ n}o, M., Luridiana, V., 
\& Castander, F.J. 2000, A\&A, 360, L5 

\bibitem[\protect\citeauthoryear{Cervi{\~ n}o et 
al.}{2003}]{Cetal03} Cervi{\~ n}o, M., Luridiana, V., P{\' e}rez, 
E., V{\'{\i}}lchez, J.M., \& Valls-Gabaud, D. 2003, A\&A, 407, 177 

\bibitem[\protect\citeauthoryear{Cervi{\~ n}o et 
al.}{2000}]{Cetal00} Cervi{\~ n}o, M., Kn{\" o}dlseder, J., 
Schaerer, D., von Ballmoos, P., \& Meynet, G. 2000, A\&A, 363, 970 

\bibitem[\protect\citeauthoryear{Cervi{\~ n}o \& 
Mas-Hesse}{1994}]{CMH94} Cervi{\~ n}o, M., \&  Mas-Hesse, J.M. 
1994, A\&A, 284, 749

\bibitem[\protect\citeauthoryear{Cervi{\~ n}o, Mas-Hesse, \& 
Kunth}{2002}]{CMHK02} Cervi{\~ n}o, M., Mas-Hesse, J.M., \& Kunth, 
D. 2002, A\&A, 392, 19 

\bibitem[\protect\citeauthoryear{Cervi{\~ n}o \& 
Valls-Gabaud}{2003}]{CVG03} Cervi{\~ n}o, M., \& Valls-Gabaud, D. 
2003, MNRAS, 338, 481 


\bibitem[\protect\citeauthoryear{Cervi{\~ n}o et 
al.}{2002}]{Cetal02} Cervi{\~ n}o, M., Valls-Gabaud, D., 
Luridiana, V., \& Mas-Hesse, J.M. 2002, A\&A, 381, 51 



\bibitem[\protect\citeauthoryear{Charbonnel et al.}{1996}]{Charetal96} Charbonnel, C., Meynet, G., Maeder, A.,  
\& Schaerer, D. 1996, A\&AS, 115, 339

\bibitem[\protect\citeauthoryear{Charbonnel et al.}{1993}]{Charetal93} Charbonnel, C., Meynet, G., Maeder, A., 
Schaeller, G., \& Schaerer, D. 1993, A\&AS, 101, 415


\bibitem[\protect\citeauthoryear{Charlot \& Bruzual}{1991}]{CB91} Charlot, S., \& Bruzual, G. 1991, ApJ, 367, 126 

\bibitem[\protect\citeauthoryear{Cid Fernandes et al.}{2004}]{Cid04} Cid Fernandes, R., Gonz\'alez Delgado, R.M., 
Schmitt, H., Storchi-Bergmann, T., Martins, L.P., P\'erez, E., Heckman, T., Leitherer, C., \& Schaerer, D.
2004, ApJ, 605, 105

\bibitem[\protect\citeauthoryear{Cid Fernandes et al.}{2001}]{Cid01} Cid Fernandes, R., Heckman, T., Schmitt, H., 
Gonz\'alez Delgado, R.M., \& Storchi-Bergmann, T. 2001, ApJ, 558, 81 


\bibitem[\protect\citeauthoryear{Fioc \& Rocca-Volmerange}{1997}]{FRV97} Fioc, M., \& Rocca-Volmerange, B. 
1997, A\&A, 326, 950


\bibitem[\protect\citeauthoryear{Gilfanov, Grimm, \& Sunyaev}{2004}]{GGS04} Gilfanov, M., Grimm, H.-J., 
\& Sunyaev, R. 2004, MNRAS, 351, 1365

\bibitem[Girardi et al.(2002)]{Gi02} Girardi, L., Bertelli, 
               G., Bressan, A., Chiosi, C., Groenewegen, M.A.T.,
               Marigo, P., Salasnich, B., \& Weiss, A.\ 2002, A\&A, 391, 195 

\bibitem[\protect\citeauthoryear{Girardi \& 
Bica}{1993}]{GB93} Girardi, L., \& Bica, E. 1993, A\&A, 274, 279 

\bibitem[\protect\citeauthoryear{Girardi et al.}{2000}]{Gietal00} Girardi, L., Bressan, A., Bertelli, G., 
\& Chiosi, C. 2000, A\&AS, 141, 371  

\bibitem[\protect\citeauthoryear{Gonz\'alez Delgado et al.}{2004}]{G04} Gonz\'alez Delgado, R.M., Cid Fernandes, R., 
P\'erez, E., Martins, L.P., Storchi-Bergmann, T., Schmitt, H., Heckman, T., \& Leitherer, C. 
2004, ApJ, 605, 127

\bibitem[\protect\citeauthoryear{Gonz\'alez Delgado, Heckman, \& Leitherer}{2001}]{GHL01} Gonz\'alez Delgado, R.M., 
Heckman, T., \& Leitherer, C. 2001, ApJ, 546, 845

\bibitem[\protect\citeauthoryear{Gonz\'alez Delgado \& Leitherer}{1999}]{GL99} Gonz\'alez Delgado, R.M., 
\& Leitherer, C. 1999, ApJS, 125, 479 (GL99)

\bibitem[\protect\citeauthoryear{Gonz\'alez Delgado, Leitherer, \& Heckman}{1999}]{GLH99} 
Gonz\'alez Delgado, R.M., Leitherer, C., \& Heckman, T. 1999, ApJS, 125, 489

\bibitem[\protect\citeauthoryear{Gorgas et al.}{1999}]{Getal99} Gorgas, J., 
Cardiel, N., Pedraz, S., \&  Gonz\'alez, J.J. 1999, A\&AS, 139, 29

\bibitem[\protect\citeauthoryear{Gorgas et al.}{1993}]{Gorgas93} Gorgas, J., 
Faber, S.M., Burstein, D., Gonz\'alez, J.J., Courteau, S., Prosser, C. 1993, ApJS, 86, 153

\bibitem[\protect\citeauthoryear{Gray \& Corbally}{1994}]{spectrum} Gray, R.O., \& Corbally, C.J. 1994, AJ, 107, 742 


\bibitem[\protect\citeauthoryear{Hauschildt \& Baron}{1999}]{HB99} Hauschildt, P. H., \& Baron, E. 1999, Journal of Computational 
and Applied Mathematics, 102, 41


%\bibitem[\protect\citeauthoryear{Hauschildt, Baron, \& Allard}{1997}]{phoenix} Hauschildt, P.H., Baron, E., 
%\& Allard, F., 1997, ApJ, 483, 390

\bibitem[\protect\citeauthoryear{Hillier \& Miller}{1998}]{Hillier} Hillier, D.J., \& Miller, D.L. 1998, ApJ, 496, 407

\bibitem[\protect\citeauthoryear{Hubeny}{1988}]{Hub88} Hubeny, I. 1988, Compt. Phys. Commun., 52, 103

\bibitem[\protect\citeauthoryear{Hubeny, Lanz, \& Jeffery}{1995}]{HLJ95} Hubeny, I., Lanz, T.,
  \& Jeffery, C.S. 1995, SYNSPEC-A Users Guide 


\bibitem[\protect\citeauthoryear{Jamet et al.}{2004}]{Lucetal04} Jamet, L., P\'erez, E., Cervi\~no, M., Stasi\'nsk, G., 
Gonz\'alez Delgado, R.M., \& V\'\i lchez, J.M. 2004, A\&A, 426, 399

\bibitem[\protect\citeauthoryear{Jim\'enez et al}{2004}]{Jietal04} Jim\'enez, R., MacDonald, J., Dunlop, J.S.,
Padoan, P., \& Peacock, J.A. 2004, MNRAS, 349, 240


\bibitem[\protect\citeauthoryear{Jones}{1997}]{Jones97} Jones, L.A. 1997, Ph.D. thesis, 
Univ. North Carolina, Chapel Hill 

\bibitem[\protect\citeauthoryear{Kauffmann et al.}{2003a}]{K03a} Kauffmann, G., Heckman, T., et al. 2003a, MNRAS, 341, 33
\bibitem[\protect\citeauthoryear{Kauffmann et al.}{2003b}]{K03b} Kauffmann, G., Heckman, T., et al. 2003b, MNRAS, 341, 54
\bibitem[\protect\citeauthoryear{Kauffmann et al.}{2003c}]{K03c} Kauffmann, G., Heckman, T., et al. 2003c, MNRAS, 346, 1055

\bibitem[\protect\citeauthoryear{Kurucz}{1991}]{Kur91} Kurucz, R.L. 1991, in Stellar Atmospheres: Beyond
Classical Limits, eds. L. Crivellari, I. Hubeny, \& D.G. Hummer (Dordrecht:Kluwer), 441

\bibitem[\protect\citeauthoryear{Kurucz}{1993}]{Kurucz} Kurucz, R.L. 1993, Kurucz CD-ROM 13, ATLAS9 Stellar Atmosphere
Programs and 2 km/s Grid (Cambridge:SAO)

\bibitem[\protect\citeauthoryear{Lan{\c c}on et
al.}{2001}]{Lanetal02} Lan{\c c}on, A., Goldader, J.D., Leitherer, 
C., \& Gonz\'alez Delgado, R.M. 2001, ApJ, 552, 150


\bibitem[\protect\citeauthoryear{Lanz \&  Hubeny}{2003}]{LH03} Lanz, T., \& Hubeny, I. 2003, ApJ, 146, 417

\bibitem[\protect\citeauthoryear{Le Borgne et al.}{2003}]{LeBetal03} Le Borgne, J.-F., Bruzual, G., et al. 2003, A\&A, 402, 433

\bibitem[\protect\citeauthoryear{Lejeune, Buser, \& Cuisiner}{1997}]{Leje} Lejeune, T., Buser, R., 
\& Cuisinier, F. 1997, A\&A, 125, 229

\bibitem[\protect\citeauthoryear{Leitherer, Fritze- v.Alvensleben, \& Huchra}{1996}]{LFvAH96} Leitherer, C., 
Fritze- v.Alvensleben, U., \& Huchra, J. 1996,
ASP. Conf. Series, 98

\bibitem[\protect\citeauthoryear{Leitherer et al.}{1999}]{SB99} Leitherer, C., Schaerer, D., et al. 1999, ApJS, 123, 3

\bibitem[\protect\citeauthoryear{Leonardi \& Rose}{2003}]{LR03} Leonardi, A.J., \& Rose, J. A. 2003, AJ, 126, 1811
 
\bibitem[\protect\citeauthoryear{Marigo \& Girardi}{2001}]{MG01} Marigo, P.~\& 
               Girardi, L.\ 2001, A\&A, 377, 132 
\bibitem[\protect\citeauthoryear{Martins et al.}{2004}]{paperI} Martins, L. Gonz\'alez Delgado, R.M., 
Leitherer, C., Cervi\~no, M., Hauschildt, P.H., \& Hubeny, I. 2004, MNRAS, submitted
               
\bibitem[\protect\citeauthoryear{Massey \& Olsen}{2003}]{MO03} Massey, P., Olsen, K.A.G. 2003, AJ, 125, 2867 

\bibitem[\protect\citeauthoryear{Mayya}{1997}]{Mayya97} Mayya, Y. D. 1997, ApJ, 482, L149

\bibitem[\protect\citeauthoryear{Meynet et al.}{1994}]{Meynet} Meynet, G., Maeder, A., Schaeller, G., Schaerer, D., 
\& Charbonnel, C. 1994, A\&A, 103, 97

\bibitem[\protect\citeauthoryear{Origlia et al.}{1999}]{Orietal99} Origlia, L., Goldader, J.D., Leitherer, C., 
Schaerer, D., \& Oliva, E. 1999, ApJ, 514, 96

\bibitem[\protect\citeauthoryear{Pauldrach, Hoffmann, \& Lennon}{2001}]{Pauldrach} Pauldrach, A.W.A, Hoffmann, T.L., 
\& Lennon, M. 2001, A\&A, 375, 161
\bibitem[\protect\citeauthoryear{Pettini et al.}{2000}]{pettini} 
Pettini, M., Steidel, C.C., Adelberger, K.L., Dickinson, M., \& Givalisco, M. 2000, ApJ, 528, 96

\bibitem[\protect\citeauthoryear{Renzini \&
Buzzoni}{1986}]{RB86} Renzini, A., \& Buzzoni, A. 1986, 
Spectral Evolution of Galaxies. eds. Chiosi C., Renzini A., Reidel, Dordrecht, 195


\bibitem[\protect\citeauthoryear{Rose}{1984}]{Rose84} Rose, J.A. 1984, AJ, 89, 1238
\bibitem[\protect\citeauthoryear{Rose}{1985}]{Rose85} Rose, J.A. 1985, AJ, 90, 1927
\bibitem[\protect\citeauthoryear{Rose}{1994}]{Rose94} Rose, J.A. 1994, AJ, 107, 206
\bibitem[\protect\citeauthoryear{Rose, Stetson, \& Tripicco}{1987}]{RST87} Rose, J.A., Stetson, P.B., \& 
Tripicco, M.J. 1987, AJ, 94, 1202

\bibitem[\protect\citeauthoryear{Salpeter}{1955}]{Sal55} Salpeter, E.E. 1955, ApJ, 121, 161

\bibitem[\protect\citeauthoryear{Schaerer et al.}{1993a}]{Schaetal93a} Schaerer, D., Charbonnel, C., 
Meynet, G., Maeder, A., \& Schaller, G. 1993a, A\&AS, 102, 339

\bibitem[\protect\citeauthoryear{Schaerer et al.}{1993b}]{Schaetal93b} Schaerer, D., Meynet, G., Maeder, A., 
\& Schaller, G. 1993b, A\&AS, 98, 523

\bibitem[\protect\citeauthoryear{Schaller et al.}{1992}]{Schalleretal92} Schaller, G., Schaerer, D., Meynet, G., 
\& Maeder, A. 1992, A\&AS, 96, 269

\bibitem[\protect\citeauthoryear{Schulz et al.}{2002}]{galev02} Schulz, J., Fritze- v.Alvensleben, U., Moller, C.S., 
\& Fricke, K.J. 2002, A\&A, 392, 1 

\bibitem[\protect\citeauthoryear{Smith, Norris, \& Crowther, P.A.}{2002}]{smith} Smith, L. J., Norris, R.P.F., 
\& Crowther, P.A. 2002, MNRAS, 337, 1309

\bibitem[\protect\citeauthoryear{Steidel et al.}{1996}]{steidel} 
Steidel, C.C., Giovalisco, M., Pettini, M., Dickinson, M., \& Adelberger, K. 1996, ApJ, 462, L17

\bibitem[\protect\citeauthoryear{Tadhunter et al.}{2004}]{Tad} Tadhunter, C., Robinson, T.G., 
Gonz\'alez Delgado, R.M., Wills, K., \& Morganti, R. 2004, MNRAS, in press (astro-ph/0410108)

\bibitem[\protect\citeauthoryear{Tinsley}{1968}]{Tin68} Tinsley, B.M. 1968, ApJ, 151, 547

\bibitem[\protect\citeauthoryear{Tinsley \&
Gunn}{1976}]{TG76} Tinsley, B.M., \& Gunn J.E. 1976, ApJ,
203, 52

\bibitem[\protect\citeauthoryear{Vazdekis}{1999}]{Vaz99} Vazdekis, A. 1999, ApJ, 513, 224 
\bibitem[\protect\citeauthoryear{Vazdekis}{2001}]{Vaz01} Vazdekis, A. 2001, ApJSS, 513, 276, 839
\bibitem[\protect\citeauthoryear{Vazdekis \& Arimoto}{1999}]{VazA99} Vazdekis, A., \& Arimoto, N.
 1999, ApJ, 525, 144

\bibitem[\protect\citeauthoryear{V\'azquez \& Leitherer}{2004}]{VL04} V\'azquez, G., \& Leitherer, C. 
(in preparation)

\bibitem[\protect\citeauthoryear{Walborn et al.}{1995}]{Watlas} Walborn, N.R., Lennon, D.J., Haser, S.M., 
Kudritzki, R-P., \& Voels, S.A. 1995, PASP, 107, 104
\bibitem[\protect\citeauthoryear{Worthey et al.}{1994}]{W94} Worthey, G., Faber, S.M., Gonz\'alez, J.J., 
Burstein, D. 1994, ApJS, 94, 687
\bibitem[\protect\citeauthoryear{Worthey \& Ottaviani}{1997}]{WO97} Worthey, G., \& Ottaviani, D.L. 1997, ApJS, 111, 377

\bibitem[\protect\citeauthoryear{Yi}{2003}]{Yi03} Yi, S.K. 2003, ApJ, 582, 202 
\end{thebibliography}
\end{document}